\documentclass{emulateapj}
\usepackage{graphicx}
\usepackage{epstopdf}
\usepackage{longtable}
\usepackage{subfigure}

\newcommand{\ltsimeq}{\raisebox{-0.6ex}{$\,\stackrel 
        {\raisebox{-.2ex}{$\textstyle <$}}{\sim}\,$}} 
\newcommand{\gtsimeq}{\raisebox{-0.6ex}{$\,\stackrel 
        {\raisebox{-.2ex}{$\textstyle >$}}{\sim}\,$}}

\newcommand{\cii}{[C\,{\sc ii}]}

\newcommand{\mgii}{Mg\,{\sc ii}}

\newcommand{\lya}{Ly$\alpha$}

\newcommand{\myemail}{chris.willott@nrc.ca}

\def\co21{CO\,(2-1)}

\shorttitle{A wide dispersion in star formation rate and dynamical mass}
\shortauthors{Willott et al.}

\begin{document}


\title{A wide dispersion in star formation rate and dynamical mass of $10^8$ solar mass black hole host galaxies at redshift 6}


\author{Chris J. Willott}
\affil{NRC Herzberg, 5071 West Saanich Rd, Victoria, BC V9E 2E7, Canada}
\email{\myemail}

\author{Jacqueline Bergeron and Alain Omont}
\affil{Sorbonne Universit\'e, UPMC Paris 6 \& CNRS, UMR 7095, Institut d'Astrophysique de Paris, 98 bis bd Arago, F-75014 Paris, France}




\begin{abstract}
ALMA [CII] line and continuum observations of five redshift $z>6$ quasars are presented. This sample was selected to probe lower black hole mass quasars than most previous studies. We find a wide dispersion in properties with CFHQS J0216$-$0455, a low-luminosity quasar with absolute magnitude $M_{1450}=-22.2$, remaining undetected implying a limit on the star formation rate in the host galaxy of $\ltsimeq 10\,M_\odot\,{\rm yr}^{-1}$, whereas other host galaxies have star formation rates up to hundreds of solar masses per year. Two other quasars have particularly interesting properties. VIMOS2911 is one of the least luminous $z>6$ quasars known with $M_{1450}=-23.1$, yet its host galaxy is experiencing a very powerful starburst. PSO\,J167$-$13 has a broad and luminous [CII] line and a neighbouring galaxy a projected distance of 5\,kpc away that is also detected in the [CII] line and continuum. Combining with similar observations from the literature, we study the ratio of [CII] line to far-infrared luminosity finding this ratio increases at high-redshift at a fixed far-infrared luminosity, likely due to lower dust content, lower metallicity and/or higher gas masses. We compile a sample of 21 high-redshift quasars with dynamical masses and investigate the relationship between black hole mass and dynamical mass. The new observations presented here reveal dynamical masses consistent with the relationship defined  by local galaxies. However, the full sample shows a very wide scatter across the black hole mass -- dynamical mass plane, whereas both the local relationship and simulations of high-redshift quasars show a much lower dispersion in dynamical mass.
\end{abstract}


\keywords{cosmology: observations --- galaxies: evolution --- galaxies: high-redshift --- quasars: general}



\section{Introduction}

Massive galaxies in the nearby universe have supermassive (greater
than a million times the mass of the Sun) black holes lurking at their
centers. In most cases these black holes are inactive as there is
little gas or stars close enough to the event horizon to be
accreted. However, the very existence of these black holes shows that
a large amount of material must have been accreted earlier in the
lifetime of the galaxy when it would have been viewed as an active
galactic nucleus (AGN), and when particularly luminous and not
obscured, as a quasar. The masses of local supermassive black holes
are correlated with the velocity dispersion of the bulges in their
host galaxies \citep{Ferrarese:2000}. This has led to almost two
decades of study into the physical reason for this correlation, in
which AGN feedback and merging are key components (see
\citeauthor{Kormendy:2013} 2013 for a review). Studying relations
between galaxies and quasar activity out to the earliest cosmic epochs
are key for understanding this puzzle.

Pioneering work with bolometer cameras showed that between 20\% and 40\%
of very high-redshift ($z>5$) quasars have detectable millimeter or
sub-millimeter emission from cool dust
\citep{Bertoldi:2003,Robson:2004,Wang:2008,Wang:2011,Omont:2013}. The relatively weak
correlation between far-infrared luminosity and quasar bolometric
luminosity, in addition to spectral energy distribution fits for a few
quasars, suggest the predominant source of energy that heats the dust
comes from extreme starbursts with star formation rates
$\gtsimeq 500 M_\odot\,{\rm yr}^{-1}$, rather than the quasar itself \citep{Wang:2011}.

This field has undergone a revolution with the superb sensitivity and
spatial resolution of the Atacama Large Millimeter Array (ALMA). The
host galaxies of almost all $z\approx 6$ quasars are now detectable
with ALMA in both continuum and the \cii\ $158\mu$m emission line,
providing information on their dust properties, chemical enrichment,
star formation rate and kinematics. The highest redshift quasar hosts
show a wide range of star formation rates
\citep{Omont:2013,Willott:2013,Venemans:2012,Venemans:2016,Wang:2016}
and the rate for moderate luminosity quasars is significantly lower at
$z=6$ than at $z=2$ \citep{Willott:2015}, implying a looser connection
between these two processes at an early epoch.

Previous studies have shown that the black hole masses of $z>5$
quasars lie above that expected for their dynamical masses determined
from the \cii\ line if they were to lie on the same tight correlation
in place at low redshift
\citep{Walter:2004,Wang:2013,Wang:2016}. Although unknown disk
inclinations can be an issue for individual sources, the full sample
should be viewed at random orientations within those allowed by AGN
unification schemes \citep{Carilli:2006,Ho:2007}. If this
observational result is true it shows that black hole accretion was
much more rapid in the early universe, possibly requiring
super-Eddington accretion or extremely massive black hole seeds
\citep{Volonteri:2015}. However, when studying the most luminous
quasars in an optically-selected sample only those with the most
massive black holes get selected and therefore if there is a wide
dispersion in the relation between black hole and dynamical mass,
there is an expectation of a bias to high black hole masses in the
observed sub-sample compared to that in a volume-limited sample
\citep{Willott:2005a,Lauer:2007}. Indeed, in a previous study of three
moderate luminosity quasars with lower black hole mass, we showed that
their dynamical masses match those of local galaxies
\citep{Willott:2015}.

In this paper we extend the sample size of our study of moderate
luminosity quasars with black holes in the
mass range $10^7<M_{\rm BH} <10^{9}M_\odot$ at $z>6$. In Section 2 we present
the new ALMA observations and data processing. Section 3 details the
results for each quasar. Section 4 compares the \cii\ line and far-IR
luminosities and studies how the ratio between these two depends on
far-IR luminosity and redshift. Section 5 analyzes the correlation
between black hole and dynamical mass in the light of these new
observations. Cosmological parameters of
$H_0=67.8~ {\rm km~s^{-1}~Mpc^{-1}}$, $\Omega_{\mathrm M}=0.307$ and
$\Omega_\Lambda=0.693$ \citep{Planck-Collaboration:2014} are assumed
throughout.

\section{Observations}

The high-redshift quasars observed in ALMA Cycle 3 project
2015.1.00606.S are CFHQS\,J021627$-$045534 (J0216$-$0455),
CFHQS\,J022122$-$080251 (J0221$-$0802), CFHQS\,J232908$-$030158
(J2329$-$0301), PSO J167.6415-13.4960 (PSO\,J167$-$13) and
VIMOS2911991793 (VIMOS2911). The quasars were selected for study with
ALMA based on their black hole masses $<10^{9}M_\odot$ to
substantially increase the number of $z>6$ quasars with such low black
hole mass observed in the \cii\ line and dust continuum. Positions,
redshifts, black hole masses and references for the observed quasars are given in Table
\ref{tab:zdata}.  J2329$-$0301 has previously been observed in ALMA
Cycle 0 project 2011.0.00243.S but was not securely detected in either
line or continuum with only a hint of \cii\ line emission at the
expected frequency \citep{Willott:2013}.  The Cycle 3 observations
presented here use more antennae to provide better sensitivity.

Observations of each quasar were made using 37 or 38 of the 12\,m
diameter antennae. The maximum baselines were between 450 and 700\,m
providing beam major axes in the range 0.6 to 0.8 arcsec for good
sensitivity to emission extended on the expected quasar host galaxy
scales of a few kpc. On-source integration times ranged from 43 to 49
minutes.  Observations of the science targets were interleaved with
those of nearby phase calibrators.

The band 6 (1.3\,mm) receivers were set to cover the frequency range
expected for the redshifted \cii\ transition
($\nu_{\rm rest}$=1900.5369 GHz) and to sample the dust
continuum. Each setup has four $\approx 2$\,GHz basebands, two pairs
of adjacent bands with a $11$\,GHz gap in between. One of these bands
was centred on the expected \cii\ location based on published \mgii\
(if available) or \lya\ redshifts. Each baseband covers a redshift
interval of $\delta z \approx 0.06$. The 128 channels in each band
have width 15.625\,MHz (equal to $17.3$\,km\,s$^{-1}$ to
$18.5$\,km\,s$^{-1}$ at these redshifts). The three bands without the
line were used to form a continuum image.

Data processing was performed with the {\small CASA} software
package\footnotemark. All five targets were imaged in a consistent
manner with no channel binning and natural baseline weighting (Briggs
robust parameter = 2). The line cube data were continuum-subtracted in
the $uv$-plane with flagging of channels showing line emission. Flux
calibration used either Pallas or a quasar from the ALMA bandpass
calibrator grid source list. The estimated flux calibration
uncertainties are 5\% and this is added in quadrature to all flux
measurement uncertainties. The continuum sensitivity achieved ranged
from 11 to 14 $\mu$Jy per beam rms. The single channel rms in the band
containing the line was between 0.09 and 0.14 mJy per beam.

\footnotetext{http://casa.nrao.edu}

\begin{figure*}
\centering
\includegraphics[width=.99\linewidth]{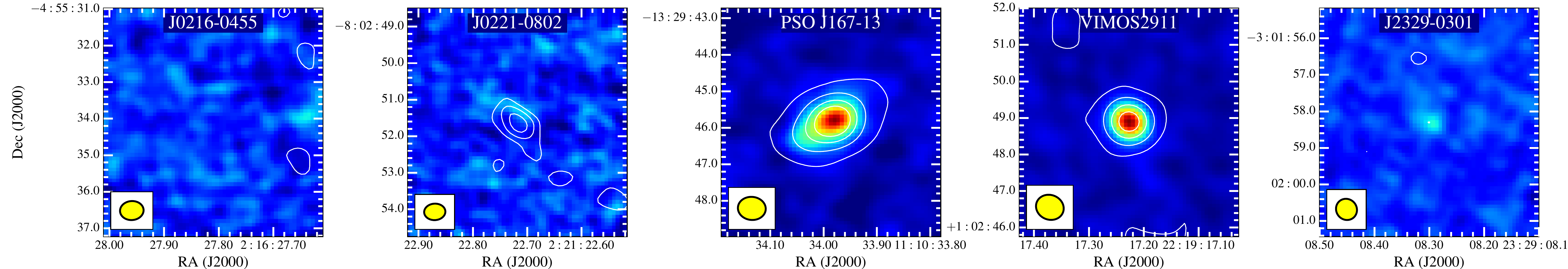}
\caption{ALMA \cii\ line (color scaling) and continuum (white contours) maps of the five $z>6$ quasars. Continuum  contours are drawn at levels of $2,3,4\,\sigma \,{\rm beam}^{-1}$ for the three CFHQS quasars and $2,10,20,30\,\sigma \,{\rm beam}^{-1}$ for PSO\,J167$-$13 and  VIMOS2911. The beam size is indicated in yellow in the lower-left corner of each panel. \cii\ is detected in three of the five quasars. Three quasars are well detected in continuum and J2329-0301 has a $2\,\sigma$ detection not well shown by the contours.}
\label{fig:contlinemaps}
\end{figure*}

Measurements from the ALMA data are given in Table
\ref{tab:almadata}. Line moment zero and continuum images are shown in
Figure \ref{fig:contlinemaps}. For obvious line-detected sources 
  the minimum and maximum channels to use in the moment zero image
  were those showing significant ($\gtsimeq 2\sigma$) emission in the individual channel
  maps. For sources without line detections the central 20 channels
($\approx 350$\,km\,s$^{-1}$) were used to make the moment zero
map. Extracted spectra of the sub-band where the \cii\ line is
expected are plotted in Figure \ref{fig:spec}. For detected sources an
elliptical aperture of $1.5 \times$ the measured source FWHM along the
major and minor axes was used to extract the spectrum. For sources
without line detections an elliptical aperture of $1.5 \times$ the
beam size was used. Uncertainties in each line channel and in the
continuum were determined by placing random apertures of equivalent
size within regions of the maps free of genuine emission and with
similar noise to the center.

\begin{table*}
\begin{center}
\caption{Position and redshift data\label{tab:zdata}}
\begin{tabular}{lllccccr}
\tableline
Quasar & RA   &    DEC  & $z_{\rm [CII]}$ &   $z_{\rm MgII}$ &   $z_{{\rm Ly}\alpha}$ & $M_{1450}$ & $M_{\rm BH} (M_\odot)^{\rm a}$  \\
\tableline
J0216$-$0455 &  02:16:27.81    &  $-$04:55:34.1$^{\rm b}$    & --                                 & --                                      & $6.01^{\rm e}$ & $-22.21$ & $2 \times 10^{7 \, \rm e}$ \\
J0221$-$0802 &  02:21:22.718  &  $-$08:02:51.62$^{\rm c}$  & --                                 & $6.161 \pm 0.014^{\rm f}$ & $6.13^{\rm g}$  &  $-24.45$ & $7 \times 10^{8 \, \rm f}$\\
PSO\,J167$-$13   &  11:10:33.986  &  $-$13:29:45.85$^{\rm d}$  & $6.5157 \pm 0.0008 $ & $6.508 \pm 0.001^{\rm h}$ & --                &   $-25.62$   & $4.9 \times 10^{8 \, \rm h}$\\
VIMOS2911      &  22:19:17.227  &     +01:02:48.88$^{\rm d}$  & $6.1492 \pm 0.0002 $  & --                                      & $6.156^{\rm i}$ &  $-23.10$ &  $5 \times 10^{7 \, \rm i}$ \\
J2329$-$0301 &  23:29:08.30    &  $-$03:01:58.3$^{\rm d}$    & $6.4164 \pm 0.0008 $ & $6.417 \pm 0.002^{\rm f}$   & $6.43^{\rm j}$   & $-25.25$ & $2.5 \times 10^{8 \, \rm f}$\\
\tableline
\end{tabular}
\end{center}
{\sc Notes.}---\\
$^{\rm a}$ Uncertainties on black hole masses are dominated by the 0.3 dex scatter in the calibration of the \mgii\ virial technique.\\
$^{\rm b}$ Optical position from discovery paper.\\
$^{\rm c}$ Millimeter continuum position from this work.\\
$^{\rm d}$ Millimeter line position from this work.\\
References: $^{\rm e}$  \cite{Willott:2009}; $^{\rm f}$  \cite{Willott:2010}; $^{\rm g}$ \cite{Willott:2010a}; $^{\rm h}$ \cite{Venemans:2015a}; $^{\rm i}$ \cite{Kashikawa:2015};  $^{\rm j}$   \cite{Willott:2007a}.\\ 
\end{table*}

\section{Notes on individual quasars}

\subsection{CFHQS J0216$-$0455}
 
J0216$-$0455 is the lowest luminosity quasar discovered in the CFHQS
with absolute magnitude $M_{1450}=-22.21$ \citep{Willott:2009}. The
redshift of $z=6.01$ is based on a strong and narrow \lya\ line, where
the broad red wing of this line identifies the object as an AGN rather
than a Lyman break galaxy. \mgii\ emission has not been detected for
this quasar, so the black hole mass of $2 \times 10^7 M_\odot$ is
instead estimated assuming that it accretes at the Eddington rate
\citep{Willott:2010}.

In the ALMA data there is no detection of either line or continuum
emission. The expected \cii\ redshift is based on the narrow \lya\
line so we do not think it likely that the true position of the \cii\
line is outside the observed band. The adjacent sub-band was searched
for a \cii\ line, but no line was seen. The non-detection in both line
and continuum is consistent with a low star-formation rate (
  $3\sigma$ limit of $\ltsimeq 10\,M_\odot\,{\rm yr}^{-1}$) in the
host galaxy of this very low black hole mass and luminosity quasar.

\begin{figure}
\centering
\begin{subfigure}
  \centering
  \includegraphics[width=.88\linewidth]{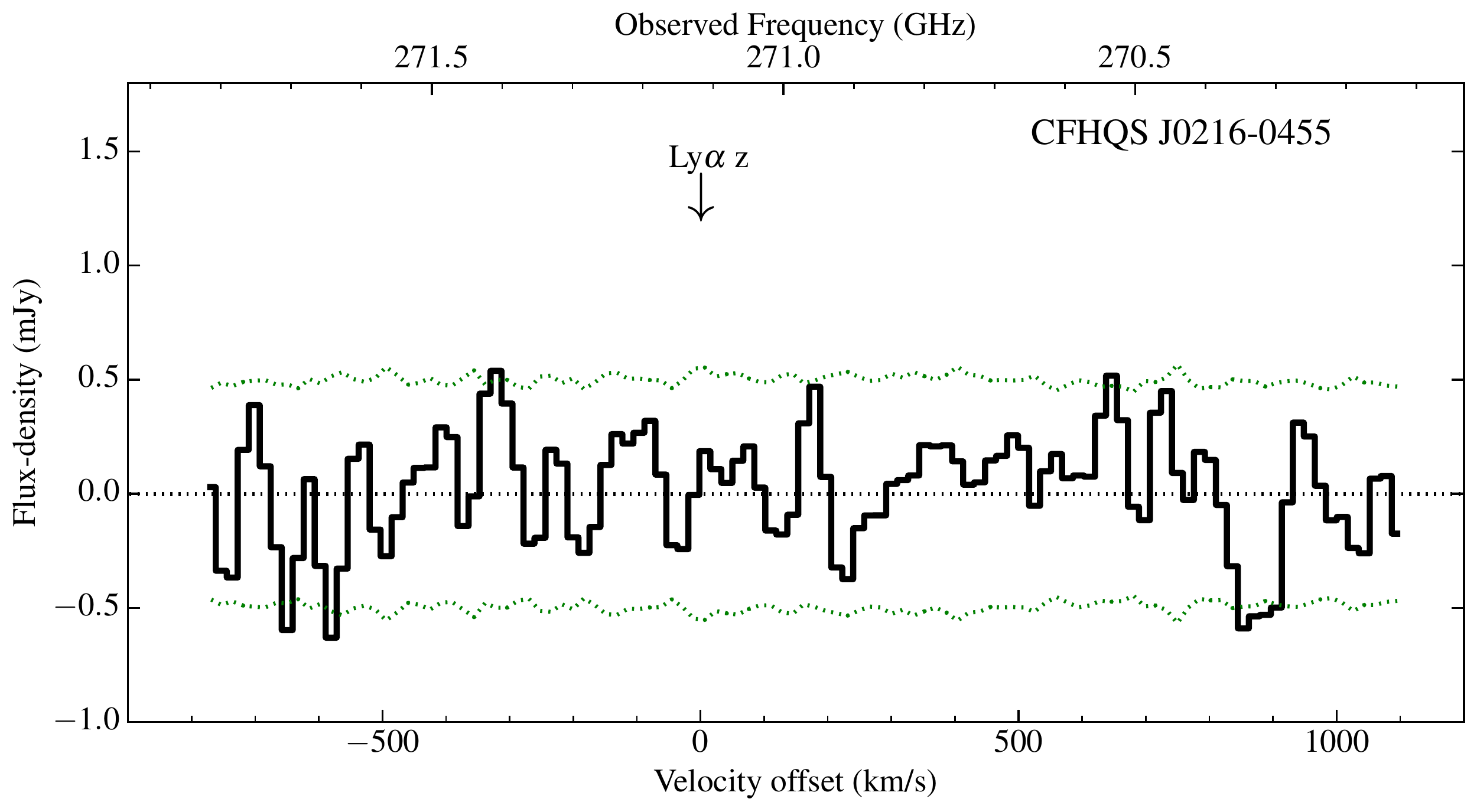}
\end{subfigure}
\begin{subfigure}
  \centering
  \includegraphics[width=.88\linewidth]{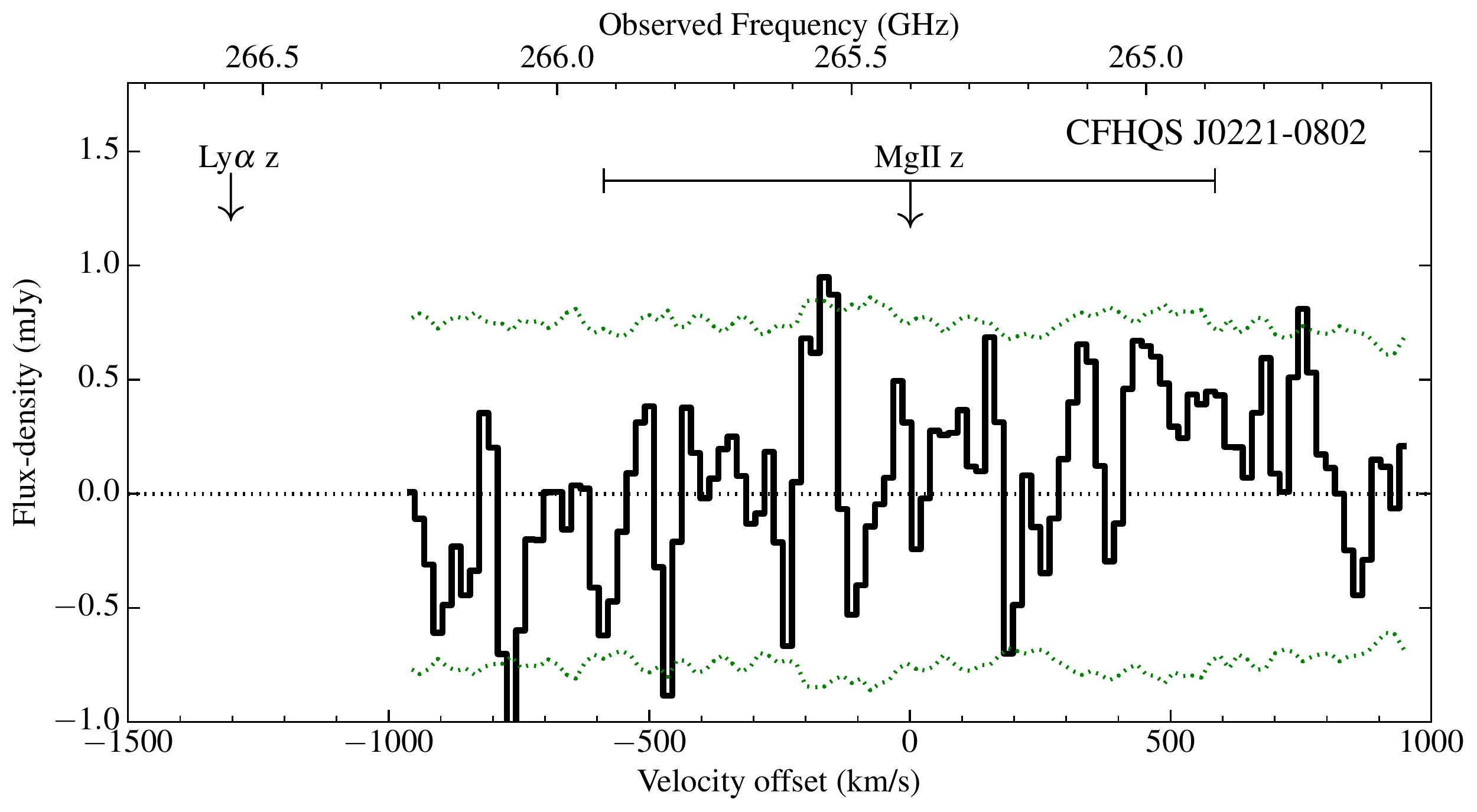}
\end{subfigure}
\begin{subfigure}
  \centering
  \includegraphics[width=.88\linewidth]{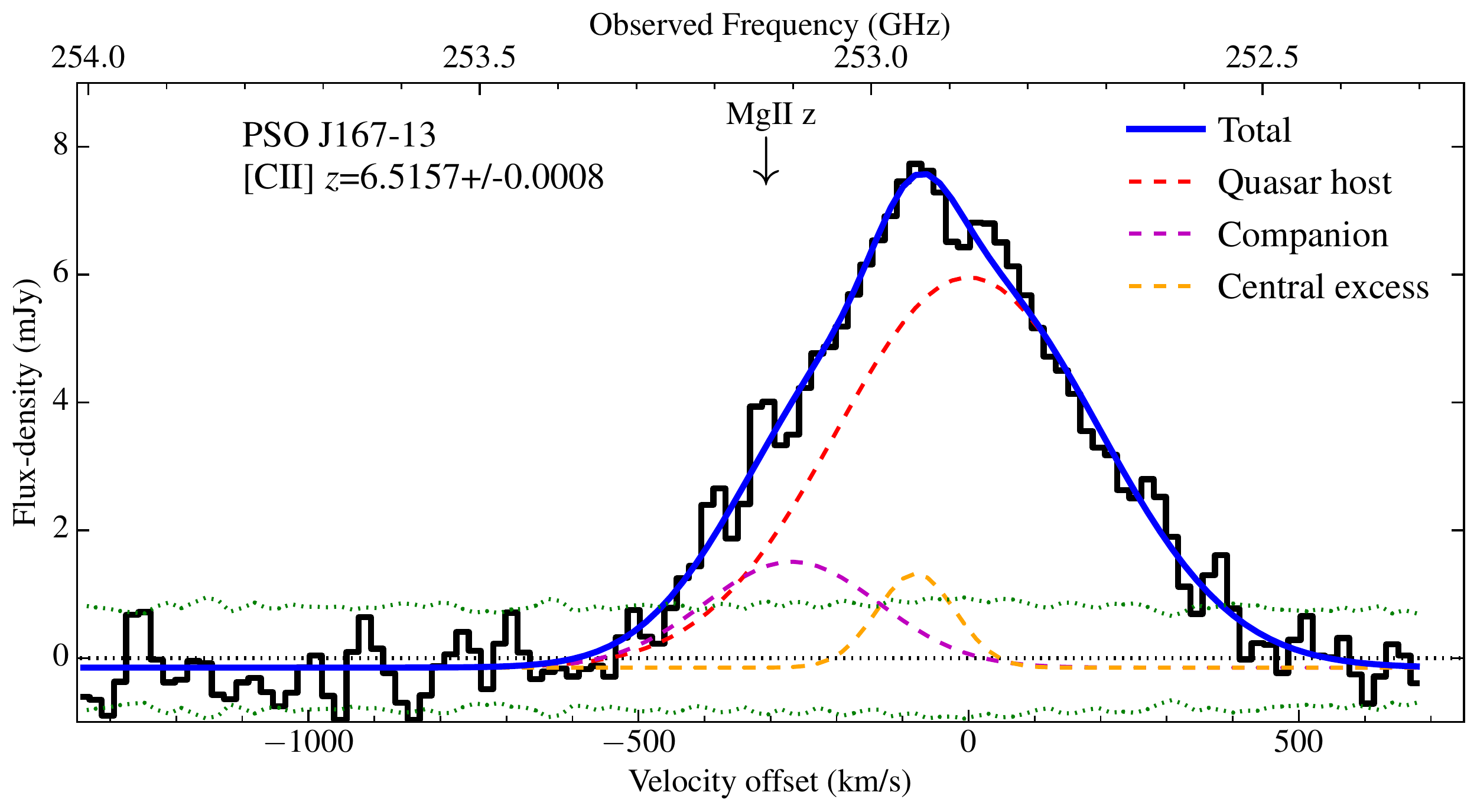}
 \end{subfigure}
\begin{subfigure}
  \centering
  \includegraphics[width=.88\linewidth]{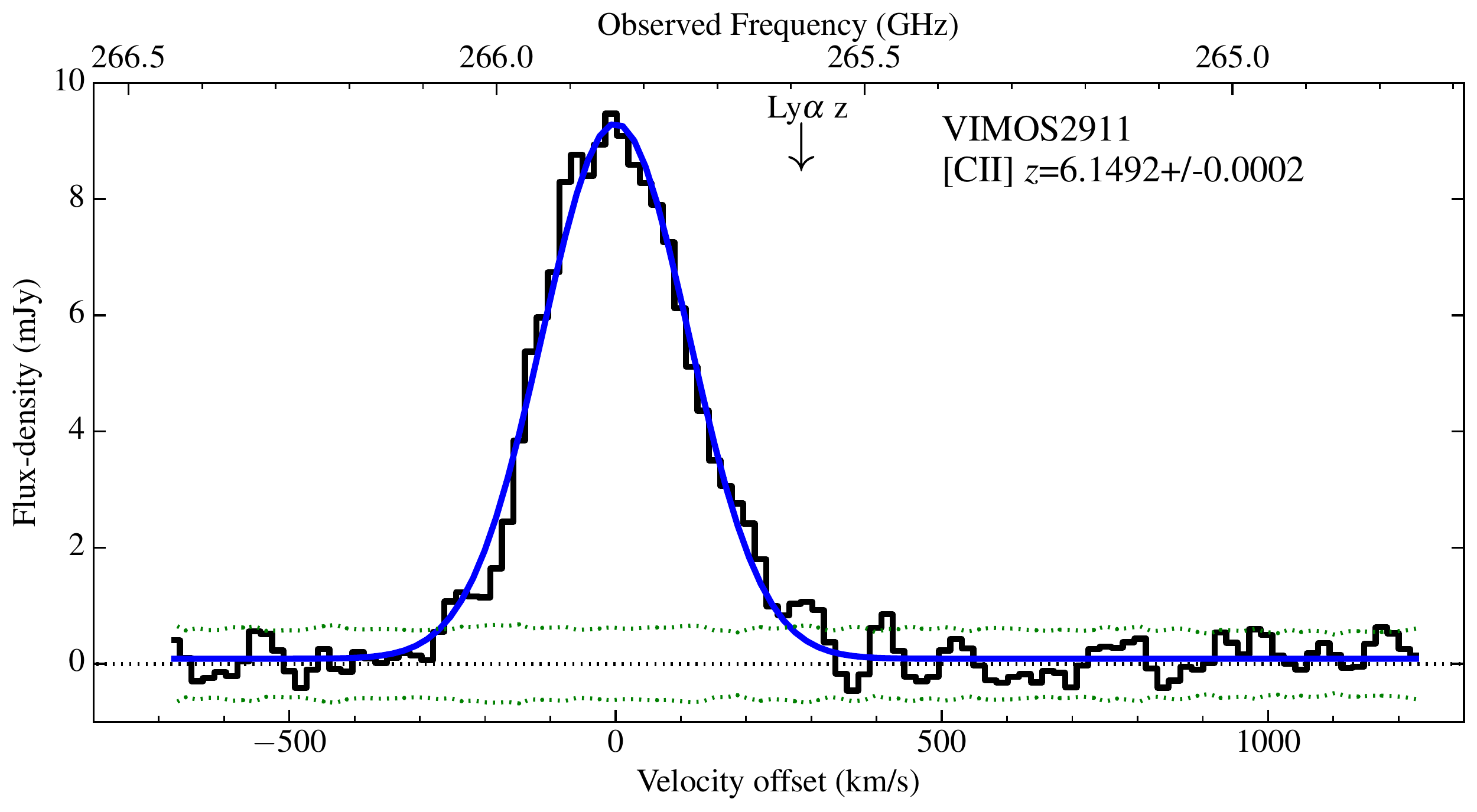}
 \end{subfigure}
\begin{subfigure}
  \centering
  \includegraphics[width=.88\linewidth]{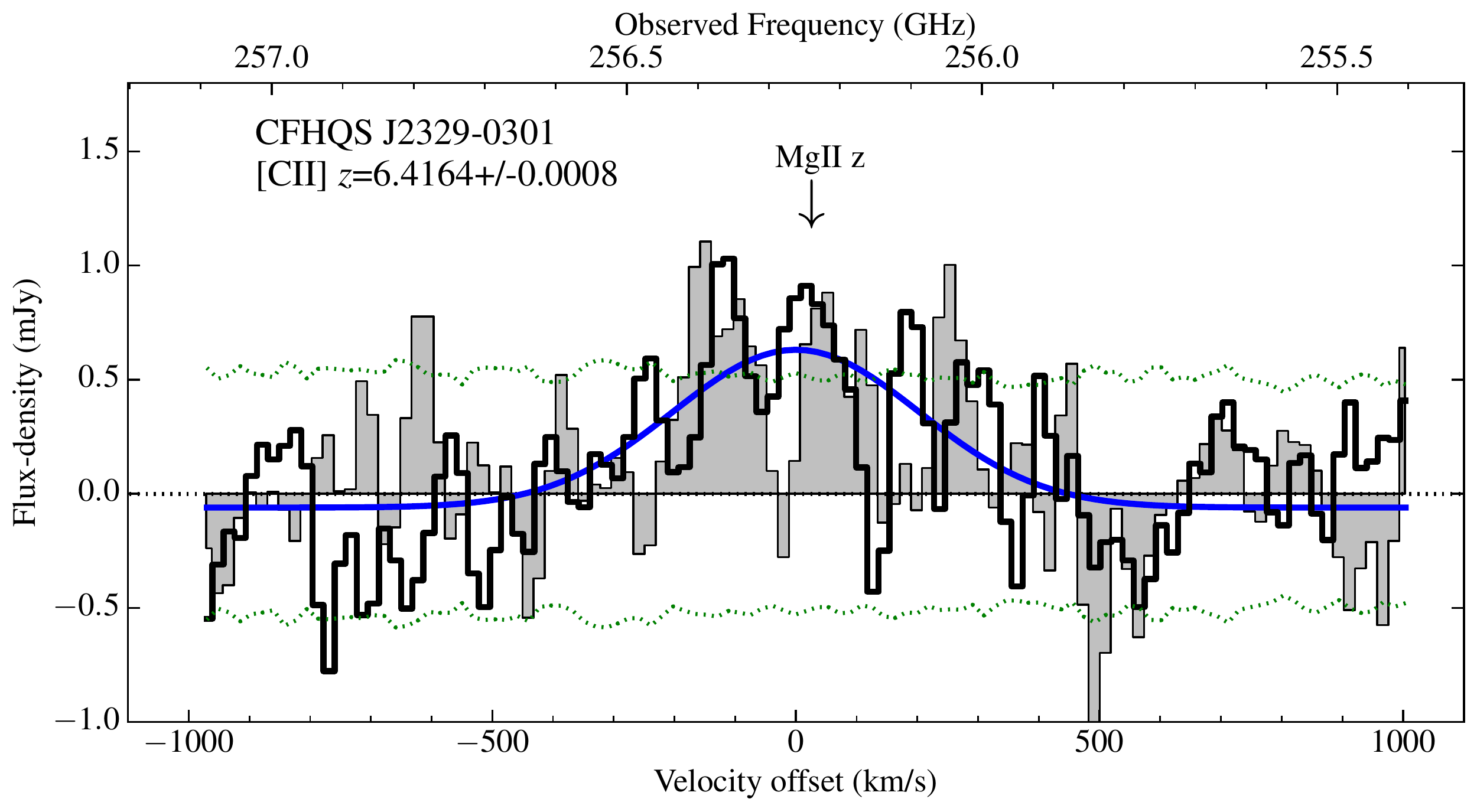}
 \end{subfigure}
\caption{ALMA band 6 spectra (black solid line) covering the expected frequency range of the \cii\ emission line. For detected lines the best-fit Gaussian model is over-plotted in blue. For PSO\,J167$-$13 the fitted curve is a combination of three Gaussians as described in the text. The $\pm 2\sigma$ noise per channel is also shown (dotted green). Expected \cii\ locations given the redshifts of rest-frame UV lines are indicated with arrows. For J2329$-$0301 the ALMA Cycle 0 spectrum is shown with gray shading.} 
\label{fig:spec}
\end{figure}

\subsection{CFHQS J0221$-$0802}
This quasar is typical of the luminosity and black hole mass of the
CFHQS survey. The \lya\ emission line is broad and asymmetric with a
narrow spike at $z=6.13$ \citep{Willott:2010a}. The near-IR spectrum
shows a relatively broad \mgii\ line that has a large uncertainty on
both its peak location ($z=6.161 \pm 0.014$) and line width, due to
the faintness of the quasar and high sky noise at this wavelength
\citep{Willott:2010}.

Figure \ref{fig:contlinemaps} shows a clear continuum detection (at
$\approx 5\,\sigma$), but no evidence for the \cii\ line is seen in
either the line map or extracted spectrum. Despite the low S/N of the
continuum detection, the source is significantly elongated at an angle
very different from the beam. The adjacent sub-band was also searched
for line emission but none was seen. For the ALMA observations we
centred the frequency sub-band based on the \mgii\ redshift. Figure
\ref{fig:spec} shows the spectrum with the \cii\ locations expected
based on the \mgii\ and \lya\ lines. The uncertainty in the \mgii\
redshift covers most of the sub-band. Unfortunately the adjacent
sub-band was set to cover lower frequency so it does not cover the
\cii\ location for the \lya\ redshift. Based on the typical \cii\ line
to continuum ratios for $z>6$ quasars with this far-IR luminosity (see
Section 4) we would expect to have detected the \cii\ line at
$>10\,\sigma$ significance. Offsets of $>1000$\,km\,s$^{-1}$ between
the \mgii\ and \cii\ redshifts have been found for other high-redshift
quasars by \cite{Venemans:2016}. Therefore we make the assumption
that the \cii\ redshift is similar to the \lya\ redshift and was
therefore missed by these observations. Table \ref{tab:almadata} gives
limits based on the non-detection of \cii\ in this quasar, but we do
not use these for the remainder of this paper because of the high
likelihood that the line was missed.

\subsection{PSO\,J167$-$13}

This is one of three $z>6.5$ quasars discovered in the Pan-STARRS1
survey by \cite{Venemans:2015a}. The optical--near-IR spectrum shows
broad \lya\ and \mgii\ giving a redshift of $6.508$ and black hole
mass of $4.9 \times 10^{8} M_\odot$, the only one in their paper with
$M_{\rm BH} < 10^9 M_\odot$, which is why we targeted it with ALMA.

The ALMA data reveal PSO\,J167$-$13 as a very luminous and extended
source in both continuum and \cii\ line emission. A Gaussian fit to
the continuum yields a deconvolved size of
$0.98 \pm 0.06 \times 0.43 \pm 0.05$ arcsec at position angle of 120$^\circ$,
equivalent to $5.5 \times 2.4$\,kpc. The continuum emission is
asymmetric being more extended to the south-east than north-west of
the peak (Figure \ref{fig:contlinemaps}).

The source is similarly extended in the \cii\ line. The peak positions
of line and continuum are co-incident and there is a roughly constant
line to continuum ratio across the whole source, however a Gaussian
fit to the \cii\ moment zero map gives a larger deconvolved size of
$1.29 \pm 0.09 \times 0.65 \pm 0.06$ arcsec at position angle of
124$^\circ$, equivalent to $7.2 \times 3.6$\,kpc. 

\begin{figure}
\centering
\begin{subfigure}
  \centering
  \includegraphics[width=1.0\linewidth]{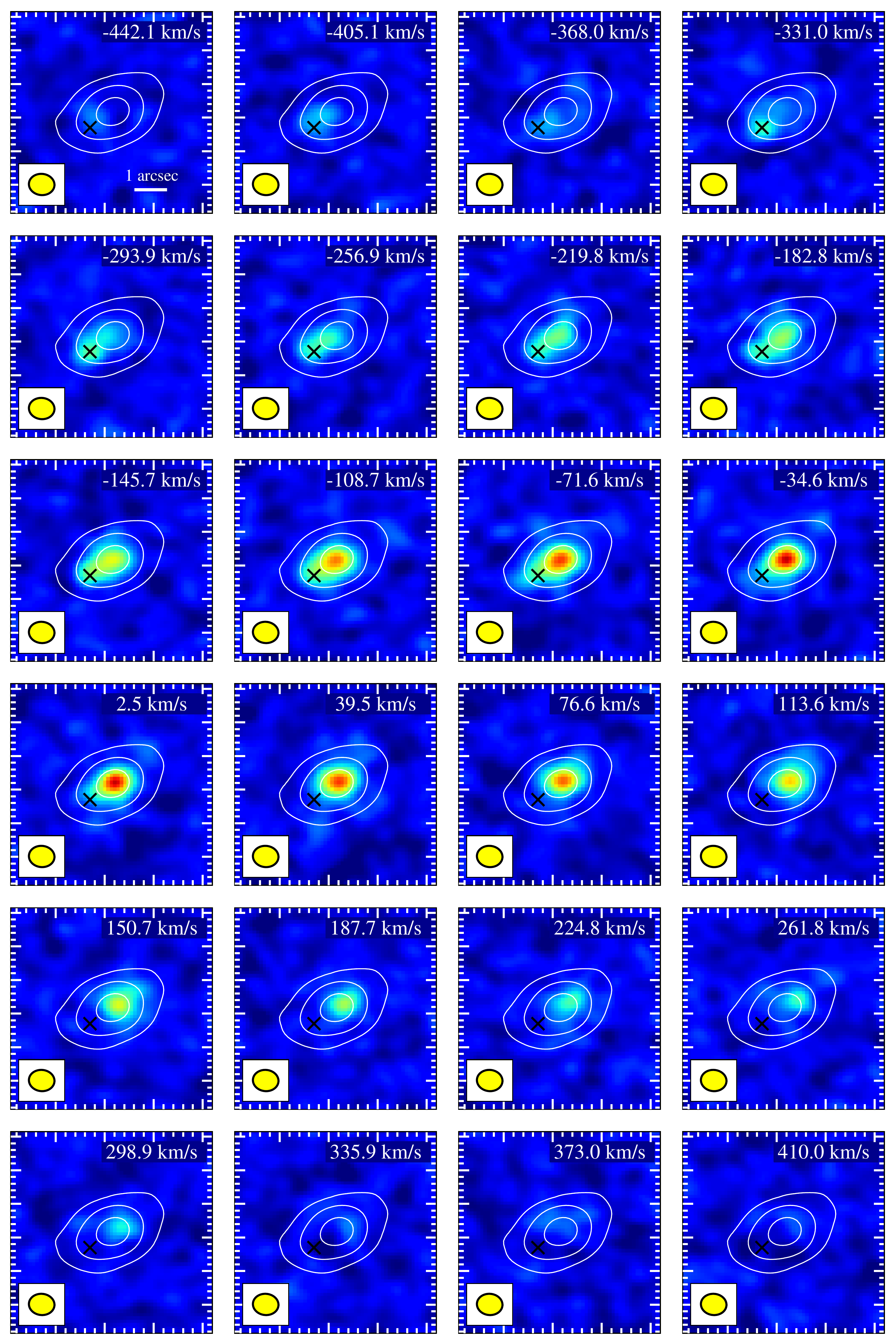}
\end{subfigure}
\caption{Channel maps from the PSO\,J167$-$13 \cii\ cube. Each image shows the \cii\ emission binned by two channels ($\delta v=37.0$\,km\,s$^{-1}$). Continuum contours at levels of $2,10,30\,\sigma \,{\rm beam}^{-1}$ are shown in white. In the velocity range $-450$ to $0$\,km\,s$^{-1}$ there is a separate source, which we call the companion galaxy, located 0.9 arcsec (projected distance 5.0\,kpc) to the south-east and marked with an x.}
\label{fig:chanmaps}
\end{figure}

To get a clearer idea of the spatial and velocity structure of the
\cii\ emission of PSO\,J167$-$13 in Figure \ref{fig:chanmaps} we show
maps in separate velocity bins. Each bin is the sum of two of the
original 128 channels, i.e. 31.25\,MHz or 37.0\,km\,s$^{-1}$. Zero
velocity is defined based on the redshift of the \cii\ peak that is
co-incident with the peak in continuum and therefore assumed to be
co-located with the quasar nucleus.  The centroid of the emission
smoothly shifts towards the north-west as velocity increases from zero
to higher values. This is reminiscent of some other $z\approx 6$
quasar host galaxies at this spatial resolution
\citep{Wang:2013,Willott:2013}, possibly indicative of a rotating
disk. As one moves to negative velocities a similar effect of shifting
the centroid to the south-east is seen, but in addition there is a
distinct source located 0.9 arcsec from the quasar centre at
11:10:34.030 $-$13:29:46.301. This source is visible at a wide range
of velocities from $-450$ to $0$\,km\,s$^{-1}$ with a peak at
$-270$\,km\,s$^{-1}$ or redshift $z=6.5090$. Because this source is
spatially and spectrally distinct from the main component we interpret
this as a companion galaxy lying at a projected distance of only
5.0\,kpc from the quasar. Similar \cii\ companions have been seen in a
few other $z>6$ quasars \citep{Decarli:2017} and galaxies
\citep{Willott:2015a,Maiolino:2015a}. Multiple UV components separated
by a few kiloparsecs are very common in the most UV-luminous galaxies
at this epoch \citep{Bowler:2017}, suggesting a high merger rate for
such galaxies. It is likely that high black hole mass quasars are hosted
in similar galaxies.

The extracted \cii\ spectrum (Figure \ref{fig:spec}) shows a broad
(500\,km\,s$^{-1}$), approximately Gaussian, profile. However, we know
from Figure \ref{fig:chanmaps} that this is composed of multiple
components. The peak in the \cii\ spectrum is offset by about
$-80$\,km\,s$^{-1}$ from the zero velocity defined by the peak at the
\cii\ quasar position. This is because of additional components at
negative velocities. In addition there is the relatively broad ($-450$
to $0$\,km\,s$^{-1}$) companion galaxy component visible at a
positional offset of 0.9 arcsec from the quasar. Therefore we fit the
\cii\ spectrum for this source with a combination of three Gaussians;
one for the quasar host, one for the companion galaxy and one for what we term the {\it central
  excess} at the peak of the integrated \cii\ profile. The relative velocities of
these Gaussian components are fixed based on the information from
Figures \ref{fig:spec} and \ref{fig:chanmaps} at $0$\,km\,s$^{-1}$ for
the quasar host, $-270$\,km\,s$^{-1}$ for the companion galaxy and
$-80$\,km\,s$^{-1}$ for the central excess and only the amplitude and
width of the Gaussians allowed to vary as free parameters. The
resulting fit is shown in Figure \ref{fig:spec} to provide a good fit
to the integrated spectrum.

In Figure \ref{fig:pvmaps} we show a position-velocity (P-V) diagram
obtained by taking a slice along the major axis defined by the quasar
host and companion. This shows clearly the two separate components in
space and velocity and suggests the companion galaxy has further
structure that would perhaps be possible to separate with higher
spatial resolution observations. The P-V diagram further suggests the
narrow central excess component at $-80$\,km\,s$^{-1}$ may be located
at 0.5 to 1.5 arcsec south-east of the quasar in roughly the same
location in projected space as the companion. This is reminiscent, but
on a scale a factor of two larger, to the \cii\ P-V diagram of the
$z=6.07$ Lyman break galaxy WMH5 \citep{Willott:2015a} that has been
interpreted as two sub-components undergoing infall along a filament
forming a massive galaxy \citep{Jones:2017} as seen in
simulations \citep{Pallottini:2017}. At positive velocities the P-V
diagram also shows a velocity gradient, as was seen in the channel
maps, indicative of a rotating disk. Follow-up at higher resolution
would be required to verify this interpretation.

For the dynamical mass estimate in Section 5 we use a
deconvolved size for the quasar host only of $1.01 \pm 0.08 \times 0.68 \pm 0.07$ arcsec
($5.6 \times 3.8$\,kpc) calculated by replacing the south-east half of
the \cii\ emission with a copy of the emission to the north-west. This
ensures only the potential galaxy disk component contributes to the
dynamical mass estimate, not that of the companion galaxy or central excess.

\cite{Venemans:2015a} quote an ionized near-zone size of $1.5 \pm 0.7$
proper Mpc for this quasar based on the \mgii\ redshift of
$z=6.508 \pm 0.001$. If one instead assumes the \cii\ redshift of
$z=6.5157 \pm 0.0008$ as the systemic redshift the near-zone size
would be $1.9 \pm 0.8$ proper Mpc. Therefore the revised redshift does
not make a significant change to the near-zone size for this quasar.

\begin{table*}
\begin{center}
\caption{Millimeter data\label{tab:almadata}}
\begin{tabular}{lccccccccc}
\tableline
Quasar              &  FWHM$_{\rm [CII]}$       & $I _{\rm [CII]}$       & $L_{\rm [CII]}$           & $f_{\rm 1.2mm}$  & $L_{\rm FIR}$           & SFR$_{\rm [CII]}$                & SFR$_{\rm FIR}$                 & $L_{\rm [CII]} / L_{\rm FIR}$ &  $M_{\rm dyn}$\\
                         &    km\,s$^{-1} $           & Jy\,km\,s$^{-1} $ & $10^8 L_\odot$     & $\mu$Jy            & $10^{11} L_\odot$    & $M_\odot\,{\rm yr}^{-1}$  & $M_\odot\,{\rm yr}^{-1}$  &   $10^{-3}$ & $M_\odot$\\
\tableline
J0216$-$0455$^a$ & --                        & $< 0.08$               & $< 0.7$             & $< 36$                 & $< 0.8$               & $< 7$                           & $<13$                              &  --                    & --  \\
J0221$-$0802$^a$ & --                        & $< 0.12$               & $< 1.2$             & $247 \pm 55$     & $ 5.7 \pm  1.3$   & $< 12$                         & $85 \pm 19$                    &  $<0.2$            & --  \\
PSO\,J167$-$13$^b$   & $469\pm 24$  & $3.84  \pm 0.20$ & $42.6 \pm 2.2$  & $1290 \pm 84$   & $27.1 \pm 1.8$  & $426 \pm 22$               & $407 \pm 26$                 & $1.57 \pm 0.13$ &  $2.3 \times 10^{11}$ \\
VIMOS2911             & $264 \pm 15$      & $2.54  \pm 0.13$ & $25.9 \pm 1.3$  & $766 \pm 47$     & $17.6 \pm 1.1$  & $259 \pm 13$               & $264 \pm 16$                 & $1.47 \pm 0.12$   & $4.4 \times 10^{10}$   \\
J2329$-$0301        & $477\pm 64$       & $0.36 \pm 0.04$  & $ 3.9 \pm 0.4$   & $39 \pm 19$       & $0.90 \pm 0.44$  & $39\pm 4$                    & $13 \pm 7$                 & $4.4 \pm 2.2$      &  $1.2 \times 10^{11}$      \\
\tableline
\end{tabular}
\end{center}
{\sc Notes.}---\\
Quoted uncertainties in $L_{\rm FIR}$,  SFR$_{\rm [CII]}$ and SFR$_{\rm FIR}$ only include measurement uncertainties, not the uncertainties in extrapolating from a monochromatic to integrated luminosity or that of the luminosity-SFR calibrations. The effect on $L_{\rm FIR}$ and SFR$_{\rm FIR}$ for dust temperatures different from the nominal $T_{\rm d} =47$\,K are given by scaling factors of 0.7 and 1.7  for $T_{\rm d} =40$\,K and 60\,K respectively. All upper limits are $3\sigma$.\\ $^a$ Limits involving \cii\ lines for these two quasars assume that the line lies within the observed ALMA frequency band. This is particularly uncertain for J0221$-$0802 where we suspect the line lies outside the observed frequency range. \\$^b$ For PSO\,J167$-$13 all data except the FWHM include the companion galaxy because it cannot be separated in the continuum. From the fit in Figure \ref{fig:spec} the quasar host component accounts for 80\% of the total \cii\ emission.
\end{table*}

\subsection{VIMOS2911}

VIMOS2911 is a faint quasar discovered at $z=6.16$ in the Subaru
High-z Quasar Survey of \cite{Kashikawa:2015}. With absolute magnitude
$M_{1450}=-23.1$ it was the second $z\approx6$ quasar, after
CFHQS\,J0216$-$0455, to be found in the rest-frame UV luminosity range
overlapping that of star-forming galaxies. The black hole mass has not
yet been measured using broad emission line widths, but assuming that
it accretes at the Eddington rate (see Section 5 for rationale of this
assumption), \cite{Kashikawa:2015} find it to have
$M_{\rm BH} = 5\times 10^7 M_\odot$.

Considering the low luminosity and black hole mass of VIMOS2911, the
ALMA observations show a surprisingly powerful starburst in the host
galaxy. There are very strong \cii\ line and continuum detections,
only a little less luminous than for PSO\,J167$-$13. In the continuum
the source has a deconvolved size of
$0.29 \pm 0.07 \times 0.08 \pm 0.14$ arcsec at position angle of
142$^\circ$, with the major axis size equivalent to $1.7$\,kpc. The
\cii\ moment zero source size is somewhat larger at
$0.52 \pm 0.05 \times 0.38 \pm 0.06$ arcsec at position angle of
20$^\circ$, equivalent to $3.0 \times 2.2$\,kpc. The greater extent of
\cii\ than continuum seen for PSO\,J167$-$13 and VIMOS2911 is
characteristic of high-redshift quasars with high S/N detections
\citep{Wang:2013}, potentially because some fraction of the dust is
heated by the quasar. However, for VIMOS2911 the low quasar luminosity
would imply a negligible contribution to the dust continuum luminosity
and its luminosity is expected to be mostly due to star formation (see
Section 4 for further discussion) .

\begin{figure}
\centering
\vspace{0.3cm}
\hspace{-0.2cm}
\begin{subfigure}
  \centering
  \includegraphics[width=.52\linewidth]{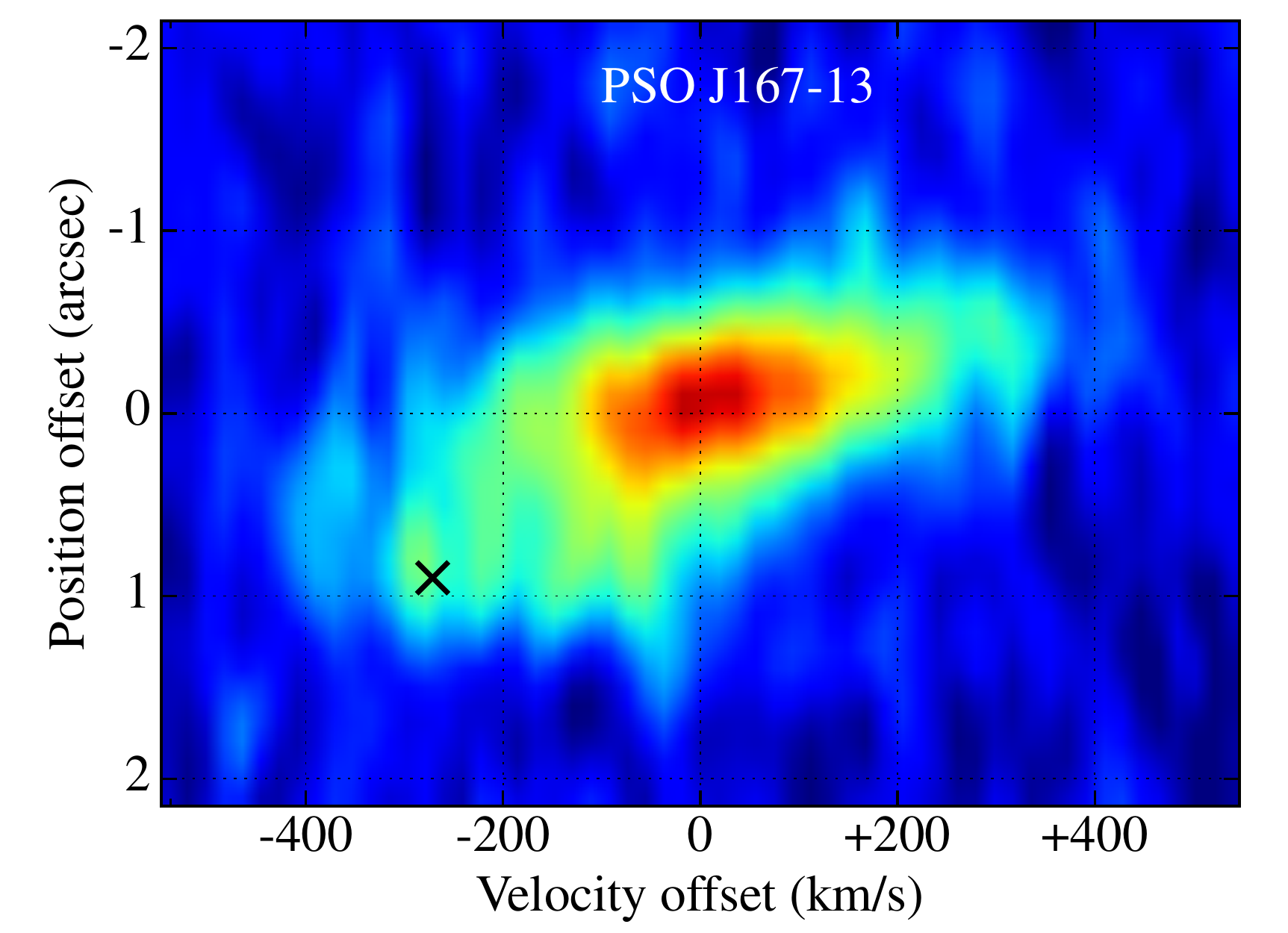}
\end{subfigure}%
\hspace{-0.6cm}
\begin{subfigure}
  \centering
  \includegraphics[width=.52\linewidth]{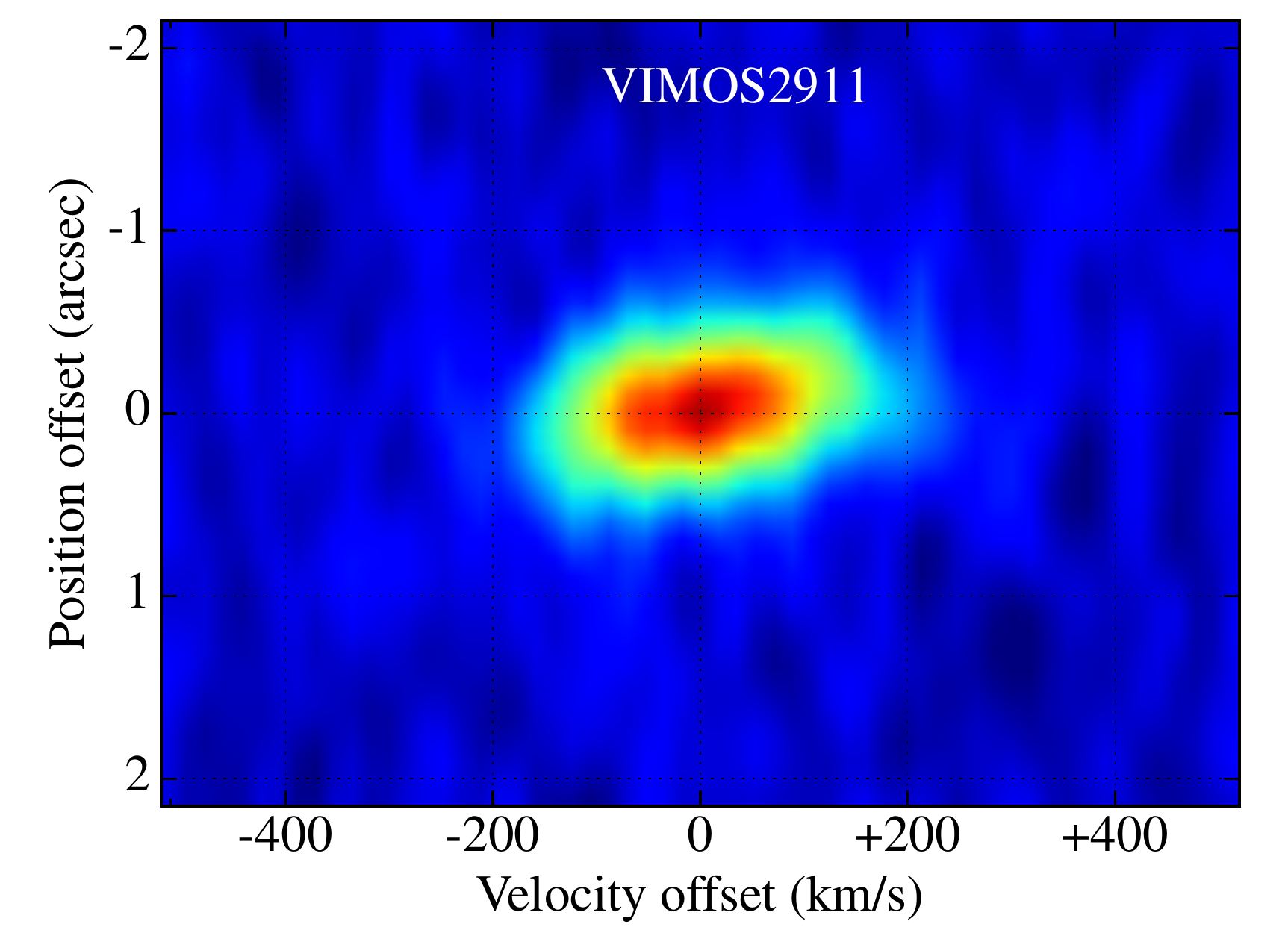}
 \end{subfigure}
\caption{Position-velocity diagrams for PSO\,J167$-$13  (left) and VIMOS2911 (right). In both diagrams the central emission is tilted showing a velocity offset from one side of the source to the other. In PSO\,J167$-$13 there is additionally extended \cii\ emitting gas corresponding to the companion galaxy, marked with an x, and the central excess at velocity $\approx  -80$\,km\,s$^{-1}$.}
\label{fig:pvmaps}
\end{figure}

The extracted spectrum of VIMOS2911 (Figure \ref{fig:spec}) is well
fit by a single Gaussian with FWHM=264\,km\,s$^{-1}$, which is fairly
narrow for a quasar host galaxy. The P-V diagram (Figure
\ref{fig:pvmaps}) shows much more symmetry than PSO\,J167$-$13 and a
gradual change in peak location versus velocity as would be found for
a marginally-resolved disk (c.f. the quasars in \citeauthor{Wang:2013} 2013).

\subsection{CFHQS J2329$-$0301}

CFHQS J2329$-$0301 is a $z=6.417$ quasar based on the \mgii\ redshift
with black hole mass $M_{\rm BH} = 2.5\times 10^8 M_\odot$. It is
embedded in a luminous \lya\ halo extending over 15\,kpc
\citep{Goto:2009,Willott:2011}. In Cycle 0 we made ALMA observations that did not
detect either line or continuum reaching limits quoted as
$I _{\rm [CII]}<0.10$\,Jy\,km\,s$^{-1} $ and $f_{\rm 1.2mm}<90 \mu$Jy
\citep{Willott:2013}. Weak \cii\ emission at a significance level
insufficient to be called a secure detection was noted in the Cycle 0
data. The new Cycle 3 data goes deeper and in the following we will
compare it with the Cycle 0 data.

Inspection of the \cii\ data cube and continuum map both show weak
emission at a position 0.5 arcsec north of the expected
position. However this expected position is based on optical data with
an astrometric uncertainty of 0.5 arcsec. As noted in
\cite{Willott:2013} there is a lower redshift galaxy 7 arcsec north of
the quasar detected in the optical and ALMA continuum. The spatial
offset between this galaxy and the weak continuum emission matches
that between this galaxy and the optical quasar thereby proving that
the \cii\ and continuum detections are indeed from the quasar.

The continuum is detected with flux-density
$f_{\rm 1.2mm}=39 \pm 19 \mu$Jy, so only $2\sigma$ and consequently
barely visible in contours on Figure \ref{fig:contlinemaps}. This low
flux explains why it was not visible in the Cycle 0 observations. The
$2\sigma$ continuum detection is taken seriously here because it
coincides with a higher significance ($10\sigma$) detection in the
\cii\ line. As seen in Figure \ref{fig:spec} the line is fairly broad
and appears to have three or more peaks, but this may be due to
noise. The redshift from a Gaussian fit almost exactly matches that of
the \mgii\ line. Also plotted on Figure \ref{fig:spec} is a thin grey
line showing the Cycle 0 spectrum. This has been extracted over the
same region as the Cycle 3 spectrum, so shows a higher flux than the
spectrum presented in \cite{Willott:2013} which used a smaller extraction
region centred 0.5 arcsec to the south. It can be seen that the Cycle
0 spectrum has positive emission at similar frequencies to the Cycle 3
spectrum, but with higher noise. The Cycle 3 flux of
$I _{\rm [CII]}=0.36 \pm 0.04$\,Jy\,km\,s$^{-1} $ is higher than
the limit in \cite{Willott:2013}. This is explained by the smaller
aperture and smaller FWHM assumed when deriving that limit.
The \cii\ moment zero deconvolved source size is
$0.85 \pm 0.33 \times 0.31 \pm 0.19$ arcsec at position angle of
25$^\circ$, equivalent to $4.8 \times 1.7$\,kpc.

\section{[CII] and FIR Luminosities}

\begin{figure}
\centering
\vspace{0.6cm}
\includegraphics[width=1.05\linewidth]{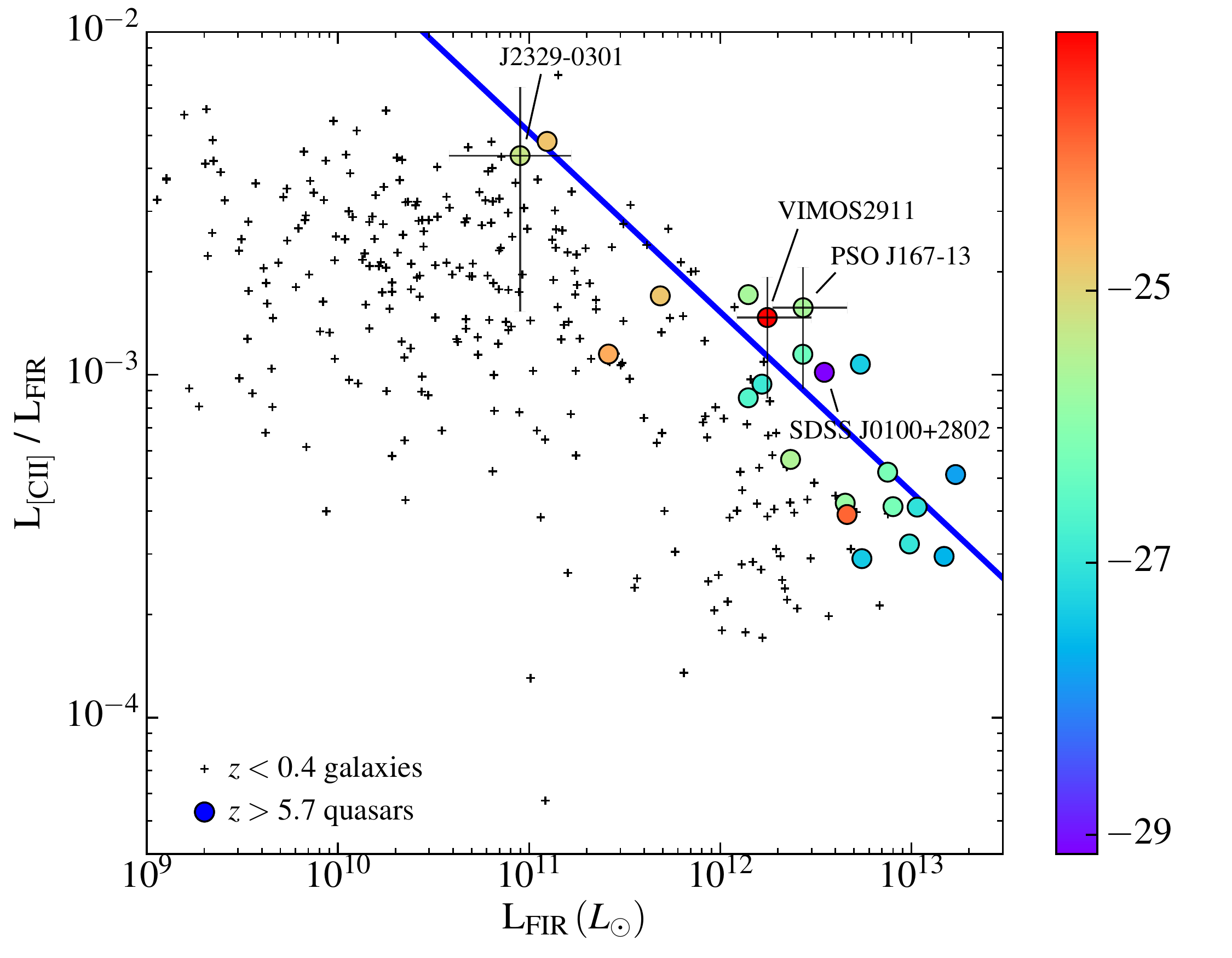}
\caption{Ratio of \cii\ to far-IR luminosity against far-IR luminosity for samples of low-redshift galaxies \citep{Gracia-Carpio:2011,Sargsyan:2012} and high-redshift quasars \citep{Maiolino:2005,Venemans:2012,Venemans:2016,Venemans:2017,Wang:2013,Wang:2016,Willott:2013,Willott:2015,Banados:2015,Decarli:2017}. The three quasars from this paper are marked by error bars and labeled. Error bars are omitted for literature sources for clarity. The $z>5.7$ quasars are color-coded by their absolute UV magnitudes (see color bar), a direct measure of AGN luminosity. The best-fit relation between the ratio and $L_{\rm FIR}$ for our 22 $z>5.7$ quasar sample is plotted with a blue line. This figure shows that (i) the high-redshift quasars follow a relation between the \cii\ to far-IR ratio and  far-IR luminosity, (ii) the relation is offset at higher redshift in the direction of stronger \cii\ lines, (iii) the far-IR luminosity is correlated with AGN luminosity but with large scatter as illustrated by the similar  $L_{\rm FIR}$ of VIMOS2911 and SDSS\,J0100+2802 (also labeled) despite the factor of $>100$ difference in AGN luminosity. }
\label{fig:ciiratio}
\end{figure}

From the observed continuum and line measurements we make
determinations of far-infrared and \cii\ line luminosities and from
these estimates of the star formation rate.  The far-IR luminosity
$L_{\rm FIR}$ integrated over rest-frame 42.5 to 122.5\,$\mu$m is
determined from the observed 1.2\,mm continuum assuming a greybody
spectrum with dust temperature, $T_{\rm d} =47$\,K and emissivity
index, $\beta=1.6$, to be consistent with previous works. There are
few measurements of $z>6$ quasar dust SEDs extending into the observed
millimeter, but \cite{Venemans:2016} measured the continuum slope in
ALMA data of one $z=6.6$ quasar and found it implied a temperature of
$37^{+11}_{-7}$\,K, consistent with our assumed temperature. We also calculate the far-IR luminosity for two other values of dust temperature to give an indication of the likely systematic uncertainty. For $T_{\rm d} =40$\,K and $T_{\rm d} =60$\,K the scaling factors to apply to $L_{\rm FIR}$ are 0.7 and 1.7 respectively. This uncertainty due to dust temperature is included in the error bars plotted in Figure \ref{fig:ciiratio}.

To
convert from $L_{\rm FIR}$ to SFR we use the relation 
\begin{equation}
{\rm SFR}\, (M_\odot\,{\rm yr}^{-1})=1.5\times10^{-10}L_{\rm FIR}\, (L_\odot)
\end{equation}
appropriate for a Chabrier IMF \citep{Carilli:2013}. Note this assumes
that all the dust is heated by star formation and not by the
quasar. The star formation rate can also be determined from the \cii\
luminosity. We use the relation 
\begin{equation}
{\rm SFR}\,  (M_\odot\,{\rm yr}^{-1}) = 1.0\times10^{-7}L_{\rm [CII]} \, (L_\odot)
\end{equation}
of \cite{Sargsyan:2014} that is valid to a precision of 50\% for a wide range of low-redshift star-forming galaxies and consistent with other low-redshift studies \citep{De-Looze:2014,Herrera-Camus:2015}.

As seen in Table 2, values of star formation derived from these two methods, \cii\
line and dust continuum, are comparable for these
quasars. The two values would be equal for the case where the ratio of
$L_{\rm [CII]}/ L_{\rm FIR}$ is $1.5\times10^{-3}$. At low redshift it
is well known that the most luminous, and hottest, infrared galaxies
have a relative deficit of \cii\ emission with ratios of a few
$\times10^{-4}$, compared to a few $\times10^{-3}$ for normal
star-forming galaxies \citep{De-Looze:2014,Diaz-Santos:2017}. A number
of factors contribute to this correlation, but the overwhelming factor
appears to be an increase in the radiation field in compact starburst
regions \citep{Diaz-Santos:2017}. Within individual galaxies there is a
clear decrease in the ratio with increasing SFR density
\citep{Smith:2017}. Figure \ref{fig:ciiratio} shows the trend of a
decreasing ratio with increasing $L_{\rm FIR}$ for a compilation of
galaxies at $z<0.4$ from \cite{Gracia-Carpio:2011} and
\cite{Sargsyan:2012}.

For high-redshift galaxies a similar trend of decreasing
$L_{\rm [CII]}/ L_{\rm FIR}$ with increasing $L_{\rm FIR}$ has been
found, however at a given $L_{\rm FIR}$ the ratio appears higher than
at low redshift \citep{Stacey:2010,Brisbin:2015}. In the pre-ALMA era
it was only possible to observe ultraluminous
($L_{\rm FIR}>10^{12}L_\odot$) galaxies at high redshifts. With ALMA detections of line
and continuum down to $L_{\rm FIR}\sim 10^{10}L_\odot$ in $z>5$
galaxies the difference between the ratios at low- and high-redshift
has been reduced \citep{Willott:2015,Willott:2015a,Capak:2015a}.

In Figure \ref{fig:ciiratio} we plot the ratio of
$L_{\rm [CII]}/ L_{\rm FIR}$ against $L_{\rm FIR}$ for 22 quasars at
$z>5.7$. There is a clear correlation here that approximately matches
the slope seen at lower redshift. We fit the relationship between
$L_{\rm [CII]}$ and $L_{\rm FIR}$ for the high-$z$ quasar sample using
orthogonal distance regression that accounts for errors in both axes. We find 
\begin{equation}
\log_{10} L_{\rm FIR}=-7.35+2.11 \log_{10} L_{\rm [CII]}
\end{equation}
indicating a steeper correlation than in our previous work (slope of
1.27 in \citeauthor{Willott:2015} 2015). The difference is due to the
increased sample size and a change in fitting method that accounts for
errors in both axes. For the $L_{\rm [CII]} / L_{\rm FIR}$ ratio line
plotted in Figure \ref{fig:ciiratio} there is a logarithmic slope of
$(1/2.11)-1=-0.53$. There are some indications from our data that the
slope flattens at $L_{\rm FIR}<10^{12}L_\odot$, just as it does for
the low-redshift galaxies.

The high-redshift best fit relationship is offset by a factor of three
compared to the low-redshift galaxies. There are several factors that
may cause this offset. \cite{Fisher:2014} suggested that high-redshift
galaxies are deficient in dust, as also seen in rest-frame UV
observations \citep{Bouwens:2014a}, and have low metallicities with
properties similar to scaled-up local dwarfs. Such galaxies typically
have $L_{\rm [CII]}/ L_{\rm FIR} \approx 3\times 10^{-3}$
\citep{Cormier:2015}, similar to the low-luminosity end of the $z>5.7$
quasar relation. \cite{Narayanan:2017} present a model where the \cii\
emission arises predominantly at the surfaces of clouds and suggest it
is the higher gas mass at early epochs that causes a higher ratio of
\cii\ to continuum emission.

\cite{Smith:2017} showed that in individual nearby galaxies there is a
correlation with both metallicity and radius, in that there is
enhanced \cii\ relative to continuum, in the more metal-poor, outer
regions of the galaxies. In addition to this they also showed a strong
anti-correlation of the ratio with star formation rate density, which
also obviously increases towards the center. It is interesting to
compare this with observations of high-redshift quasars. Whilst the
quasars are mostly only just resolved in current ALMA observations, in
cases where deconvolved sizes can be measured it has always been found
that the \cii\ emission is somewhat more extended than the dust
continuum as we found for PSOJ167$-$13 and VIMOS2911 here
\citep{Wang:2013,Cicone:2015,Venemans:2016}.  This is in agreement
with the spatially-resolved results at lower redshift.

One possibility to explain the more extended line than continuum
emission in quasars is that a large fraction of the energy that heats
the dust comes from the quasar nucleus.  In Figure \ref{fig:ciiratio}
we color-coded the $z>5.7$ quasar sample with their absolute
magnitudes. A positive correlation between quasar luminosity and
far-IR luminosity does exist as has been seen previously
\citep{Wang:2011,Omont:2013,Venemans:2016}. However, there is a large amount
of scatter and significantly the most luminous and least luminous
quasars in the sample, SDSS\,J0100+2802 \citep{Wang:2016}, and
VIMOS2911 (this paper), have very similar values of both
$L_{\rm [CII]}$ and $L_{\rm FIR}$. In particular, the fact that the
most luminous known $z>6$ quasar has a relatively high
$L_{\rm [CII]}/ L_{\rm FIR}$ ratio suggests that it is not necessarily
the presence of a luminous AGN that depresses the global \cii\ to FIR ratio.

An alternative to using the ratio of \cii\ to far-IR luminosities is to calculate the rest-frame equivalent width, EW$_{\rm [CII]}$, of the \cii\ line. This quantity has the advantage of not depending on the extrapolation from a single far-IR continuum point to the integrated far-IR luminosity. For our three quasars with detections in line and continuum we find EW$_{\rm [CII]}$=1.56 (PSO\,J167$-$13), 1.74 (VIMOS2911) and 4.86 (J2329$-$0301) $\mu$m. These are comparable to the typical value of EW$_{\rm [CII]}$=1.27 $\mu$m for local starburst galaxies \citep{Venemans:2016}.

\section{Dynamical Mass and Black hole mass}

The aim of this observing program was to increase the number of
quasars at $z>6$ and black hole masses $M_{\rm BH}<10^{9}M_\odot$ with
ALMA line and continuum measurements. This region of parameter space
is sparsely populated to date with the majority of ALMA observations
of $z>6$ quasars having black hole masses $M_{\rm
  BH}>10^{9}M_\odot$. As described earlier, only by fully sampling a
wide range of black hole masses can we avoid the biases imposed by
scatter.

Measuring dynamical masses at these high redshifts with marginally
resolved data is a challenge as clear evidence for ordered kinematics
is not usually available. However, many $z>6$ quasars do show
spatially-resolved velocities with a velocity gradient from one side
of the major axis to the other
\citep{Wang:2013,Willott:2013,Venemans:2016,Decarli:2017}. Here we
follow the method of \cite{Wang:2013} in assuming that the observed
\cii\ emission comes from a thin disk where the observed velocity
structure is entirely due to rotation. The fact that most of the
spectra are not flat-topped or double-horned velocity profiles shows
that the thin disk assumption is indeed only an approximation of the
dynamical mass. The inclination angle, $i$, is determined by the ratio
of the deconvolved minor ($a_{\rm min}$) and major ($a_{\rm maj}$) axes,
$i=\cos^{-1}( a_{\rm min}/a_{\rm maj})$. The circular velocity is
$v_{\rm cir}=0.75 \,$FWHM$_{\rm [CII]} / \sin i$. Then the dynamical
mass is given by
\begin{equation}
M_{\rm dyn} = 1.16\times 10^5\, v_{\rm cir}^2 \,D \,M_\odot
\end{equation}
where $D$ is the disk diameter in kpc, calculated as $1.5 \times$ the
deconvolved Gaussian spatial FWHM.

\begin{figure}
\centering
\includegraphics[width=1.02\linewidth]{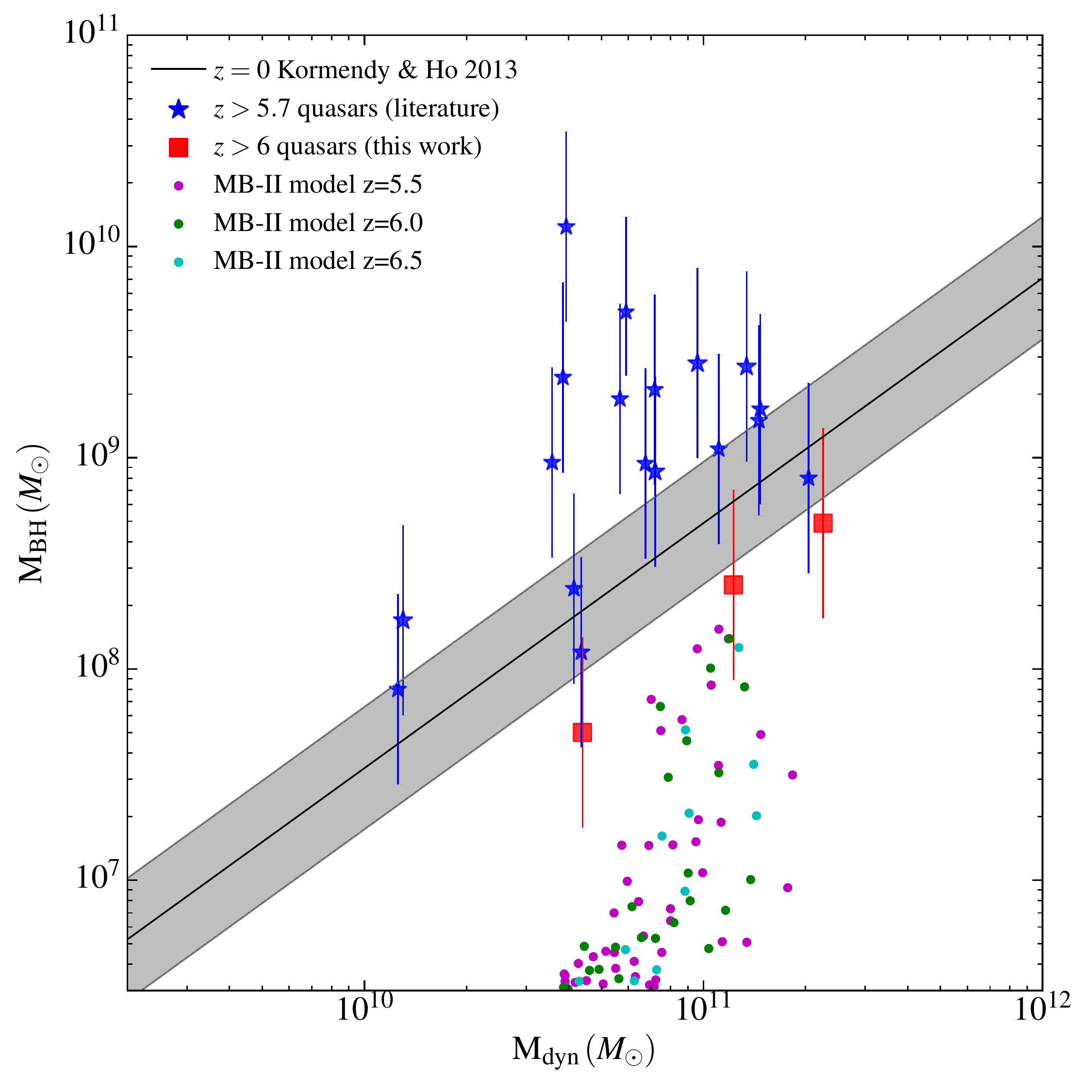}
\caption{Black hole mass versus host galaxy dynamical mass for
  $z\approx 6$ quasars. Previous observations are shown as blue stars
  and the three from this work as red squares. Error bars are not
  shown on dynamical masses because there are multiple uncertainties
  including unknown inclinations for nine quasars. Small circles show
  sub-halo masses from the MassiveBlack-II simulation of
  \cite{Khandai:2015} at three redshifts near $z=6$. The black line
  with gray shading is the local correlation $\pm 1\sigma$ scatter
  \citep{Kormendy:2013}. The new data lie below, but very close to,
  the $z=0$ relation. The full quasar sample has a wider range of
  $M_{\rm dyn}$ at a given $M_{\rm BH}$ than the simulations.}
\label{fig:mbhmdyn}
\end{figure}

Our sample consists of 21 of the 22 quasars discussed in Section 4
and plotted on Figure \ref{fig:ciiratio}.  We exclude SDSS
J1044$-$0125 because the \cii\ and CO ($6-5$) spectra differ,
indicating complex dynamics \citep{Wang:2013}. All quasars have
measured \cii\ FWHM, although for SDSS\,J1148+5251 we instead use the
CO observations of \cite{Stefan:2015} for the reasons discussed in
\cite{Wang:2016}. Four quasar host galaxies are unresolved in the
rest-frame FIR line data so for these we assume a size of
$D=4.5$\,kpc, which is the median of the resolved sources. This is a
reasonable assumption as the upper limits are generally greater than
this value. Nine quasars do not have deconvolved minor axis sizes
available, so for these we assume the random orientation angle of
55$^{\circ}$. For the three quasars presented in this paper we
determine dynamical masses of PSO\,J167$-$13
$M_{\rm dyn}=2.3 \times 10^{11} M_\odot$, VIMOS2911
$M_{\rm dyn}=4.4 \times 10^{10} M_\odot$ and J2329$-$0301
$M_{\rm dyn}=1.2 \times 10^{11} M_\odot$.

Black hole masses of the quasars are mostly based on the \mgii\ broad
emission line virial calibration as reported in
\cite{Willott:2010,De-Rosa:2011,De-Rosa:2014,Wang:2013,Venemans:2013,Venemans:2015a,Wu:2015,Shao:2017}. The
uncertainties on black hole masses are all dominated by the 0.3 dex
scatter in the calibration of the \mgii\ virial technique. For three
quasars, \mgii\ information is not available, so their black hole
masses are estimated from their UV luminosities assuming they accrete
at the Eddington luminosity as is commonly observed for such
high-redshift quasars \citep{Jiang:2007,Willott:2010,De-Rosa:2011}.

In Figure \ref{fig:mbhmdyn} we plot the black hole mass against the
dynamical mass for these 21 $z>5.7$ quasars. The three new additions
from this paper are shown as red squares. It is clear that our effort
to target quasars with low black hole mass reveals a different picture
to previous works based mostly on the most massive black holes. All
three of our new observations show black hole masses at a given
dynamical mass below, but consistent with, the $z=0$ relation of
\cite{Kormendy:2013} (equating $M_{\rm dyn}$ to $M_{\rm bulge}$). The
distribution of dynamical masses for these three quasars is consistent
with the same distribution as for quasars with black hole masses a
factor of 10 to 100 times greater. The new observations serve to
enforce the results of previous works \citep{Willott:2015,Wang:2016}
that there is no strong correlation between the two properties in
$z\sim6$ quasars, but that the scatter is very much larger in the
early universe than it is today. 

We remind the reader that all the quasar data plotted in Figure
  \ref{fig:mbhmdyn} has been calculated using the same rotating disk
  assumption for the dynamical mass, but even if some quasar hosts are
  dispersion-dominated it would not strongly affect these results.
Even with the new data it is still not yet possible to determine the
true relationship at $z=6$ between dynamical mass and black hole mass
in a volume-limited sample as there will be many black holes with
still lower mass that cannot be identified as AGN with current
observations.

High-resolution cosmological hydrodynamical simulations
can now simulate large enough volumes to trace the early growth of
supermassive black holes. Whilst high-resolution volumes containing
the most massive black holes are still out of reach, recent works have
enabled volumes containing black holes of mass up to
$M_{\rm BH} = 10^8 M_\odot$. The MassiveBlack-II simulation of
\cite{Khandai:2015} includes star formation, black hole accretion and
feedback and provides a theoretical comparison to our observations. In
Figure \ref{fig:mbhmdyn} we show the black hole mass and dynamical
mass of the dark matter halo catalogues from MassiveBlack-II at three
redshifts: $z=5.5, 6.0, 6.5$. For the dynamical mass we use the
circular velocities of the halos from the simulation and an assumed
size equal to the median of our 21 quasars ($D=4.5$\,kpc). This
enables a direct comparison of the mass within the same radius rather
than the full extent of the halo. The dynamical masses are then
calculated with Equation (4) in the same way as for the quasar
sample. 

The simulated quasars with $M_{\rm BH} \approx 10^8 M_\odot$ have
dynamical masses of $M_{\rm dyn} \approx 10^{11} M_\odot$, consistent
with some of our quasars such as J2329$-$0301, but on average somewhat
higher than the typical dynamical mass for these quasars of a few
times $10^{10} M_\odot$. The observations show a much greater scatter
than the simulations. In part this is because of a wider range of $D$
for the observations than the single value used with the simulation
data, but the observations also show a larger range of $v_{\rm
  cir}$. In particular the simulations predict a very steep
relationship with a wide range of black hole masses hosted by a narrow
range of halo masses, whereas the observations show a wide spread in
both properties. The slope of the correlation in the simulations is
much steeper than that at $z=0$ as rapid exponential accretion
causes an increase in black hole mass much faster than the slow
accumulation in dark matter halos.

\section{Conclusions}

The ALMA observations presented here for five $z>6$ quasars with black
hole masses $<10^9 M_\odot$ show a wide range of properties. The
lowest optical luminosity quasar, J0216$-$0455, is undetected in both
the \cii\ line and the continuum, indicating a star formation rate in
its host galaxy of $\ltsimeq 10\,M_\odot\,{\rm yr}^{-1}$.
J2329$-$0301, undetected in ALMA Cycle 0 data, is now just detected
with SFR $\approx 10-40\,M_\odot\,{\rm yr}^{-1}$. J0221$-$0802 has a
dust continuum detection revealing SFR
$\approx 90 \,M_\odot\,{\rm yr}^{-1}$, typical of highly star-forming
galaxies. For J0221$-$0802 we speculate the lack of a \cii\ detection
is due to a velocity offset between the line and that targeted by
these observations. VIMOS2911, despite being a low luminosity quasar,
resides in a host galaxy undergoing a powerful starburst with
estimated star formation rate of
$\approx 250 \, M_\odot\,{\rm yr}^{-1}$. The source is extended in
both line and continuum with deconvolved sizes of 2 to 3\,kpc and
there is a velocity gradient across the source.

For PSO\,J167$-$13 we found a very luminous and extended source in
both line and continuum, revealing a powerful starburst with SFR
$\approx 400 \, M_\odot\,{\rm yr}^{-1}$. In addition to an extended
(5\,kpc) disk with a velocity gradient, the \cii\ data reveal a
companion galaxy at a projected distance of 5\,kpc and probably at least one
more velocity component offset from the main galaxy spatial location.
This system appears to be an example of a high-density region where a
massive galaxy is forming by accretion of star-forming sub-clumps as
seen in some other $z>6$ galaxy observations
\citep{Willott:2015a,Maiolino:2015a,Decarli:2017,Bowler:2017} and in
simulations of \cii\ emission in the early universe
\citep{Katz:2017,Pallottini:2017}.

We have investigated the ratio of \cii\ to far-IR emission in a sample
of 22 $z>5.7$ quasars. We found a correlation between this ratio and
the far-IR luminosity with a logarithmic slope of $-0.53$, similar to
the slope seen in nearby luminous and ultraluminous galaxies. However, the
high-redshift quasar sample is offset by a factor of three compared to
the low-redshift galaxies. This increase in \cii\ line luminosity at
higher redshift could be due to lower dust content, lower metallicity
or higher gas masses. For two of our quasars where it could be
measured, the size of the \cii-emitting region is larger than the size
of the dust-emitting region, which has been seen in other high- and
low-redshift galaxies
\citep{Wang:2013,Cicone:2015,Venemans:2016,Smith:2017}.

Our study significantly increases the number of high-redshift quasars
containing moderate mass black holes with measured dynamical masses.
All three quasars in this paper fall below the
$M_{\rm BH} - M_{\rm dyn}$ relation for low-redshift galaxies,
contrary to what has been observed to date for high-redshift quasars
with higher black hole masses. The distribution of dynamical masses
for these three low black hole mass quasars is consistent with the
distribution for quasars with black hole masses ten to a hundred times
greater. There is no significant correlation between these two
properties in high-redshift quasars, but there is a much larger
scatter than seen in galaxies in the local universe. We compared our
measurements with the results of simulations of
$M_{\rm BH} \approx 10^8 M_\odot$ quasar hosts
\citep{Khandai:2015}. The simulated quasars reside in a narrower range
of halo masses than the observed quasars, with a typical value a
little higher than the median of the observations in the same black
hole mass range. However the simulations do predict a relatively weak
correlation between these two quantities at $z\approx6$, as
observed. The weak correlation is likely an effect of the rapid
accretion growth of supermassive black holes at early times. In
contrast, at later times, merging and AGN feedback tighten the
relationship.
  

\acknowledgments

Thanks to staff at the North America ALMA Regional Center for initial
processing of the ALMA data.  We thank the anonymous referee for
useful comments. This paper makes use of the following ALMA data:
ADS/JAO.ALMA\#2015.1.00606.S. ALMA is a partnership of ESO
(representing its member states), NSF (USA) and NINS (Japan), together
with NRC (Canada) and NSC and ASIAA (Taiwan), in cooperation with the
Republic of Chile. The Joint ALMA Observatory is operated by ESO,
AUI/NRAO and NAOJ. The National Radio Astronomy Observatory is a
facility of the National Science Foundation operated under cooperative
agreement by Associated Universities, Inc.



{\it Facility:} \facility{ALMA}.



\bibliography{willott}

\begin{thebibliography}{62}
\providecommand\natexlab[1]{#1}
\providecommand\JournalTitle[1]{#1}

\bibitem[{{Ba{\~n}ados} {et~al.}(2015){Ba{\~n}ados}, {Decarli}, {Walter},
  {Venemans}, {Farina}, \& {Fan}}]{Banados:2015}
{Ba{\~n}ados}, E., {Decarli}, R., {Walter}, F., {et~al.} 2015,
  \href{http://dx.doi.org/10.1088/2041-8205/805/1/L8}{\JournalTitle{\apjl},
  805, L8}

\bibitem[{{Bertoldi} {et~al.}(2003){Bertoldi}, {Carilli}, {Cox}, {Fan},
  {Strauss}, {Beelen}, {Omont}, \& {Zylka}}]{Bertoldi:2003}
{Bertoldi}, F., {Carilli}, C.~L., {Cox}, P., {et~al.} 2003,
  \href{http://dx.doi.org/10.1051/0004-6361:20030710}{\JournalTitle{\aap}, 406,
  L55}

\bibitem[{{Bouwens} {et~al.}(2014){Bouwens}, {Illingworth}, {Oesch},
  {Labb{\'e}}, {van Dokkum}, {Trenti}, {Franx}, {Smit}, {Gonzalez}, \&
  {Magee}}]{Bouwens:2014a}
{Bouwens}, R.~J., {Illingworth}, G.~D., {Oesch}, P.~A., {et~al.} 2014,
  \href{http://dx.doi.org/10.1088/0004-637X/793/2/115}{\JournalTitle{\apj},
  793, 115}

\bibitem[{{Bowler} {et~al.}(2017){Bowler}, {Dunlop}, {McLure}, \&
  {McLeod}}]{Bowler:2017}
{Bowler}, R.~A.~A., {Dunlop}, J.~S., {McLure}, R.~J., \& {McLeod}, D.~J. 2017,
  \href{http://dx.doi.org/10.1093/mnras/stw3296}{\JournalTitle{\mnras}, 466,
  3612}

\bibitem[{{Brisbin} {et~al.}(2015){Brisbin}, {Ferkinhoff}, {Nikola},
  {Parshley}, {Stacey}, {Spoon}, {Hailey-Dunsheath}, \& {Verma}}]{Brisbin:2015}
{Brisbin}, D., {Ferkinhoff}, C., {Nikola}, T., {et~al.} 2015,
  \href{http://dx.doi.org/10.1088/0004-637X/799/1/13}{\JournalTitle{\apj}, 799,
  13}

\bibitem[{{Capak} {et~al.}(2015){Capak}, {Carilli}, {Jones}, {Casey},
  {Riechers}, {Sheth}, {Carollo}, {Ilbert}, {Karim}, {Lefevre}, {Lilly},
  {Scoville}, {Smolcic}, \& {Yan}}]{Capak:2015a}
{Capak}, P.~L., {Carilli}, C., {Jones}, G., {et~al.} 2015,
  \href{http://dx.doi.org/10.1038/nature14500}{\JournalTitle{\nat}, 522, 455}

\bibitem[{{Carilli} \& {Walter}(2013)}]{Carilli:2013}
{Carilli}, C.~L., \& {Walter}, F. 2013,
  \href{http://dx.doi.org/10.1146/annurev-astro-082812-140953}{\JournalTitle{\araa},
  51, 105}

\bibitem[{{Carilli} \& {Wang}(2006)}]{Carilli:2006}
{Carilli}, C.~L., \& {Wang}, R. 2006,
  \href{http://dx.doi.org/10.1086/503872}{\JournalTitle{\aj}, 131, 2763}

\bibitem[{{Cicone} {et~al.}(2015){Cicone}, {Maiolino}, {Gallerani}, {Neri},
  {Ferrara}, {Sturm}, {Fiore}, {Piconcelli}, \& {Feruglio}}]{Cicone:2015}
{Cicone}, C., {Maiolino}, R., {Gallerani}, S., {et~al.} 2015,
  \href{http://dx.doi.org/10.1051/0004-6361/201424980}{\JournalTitle{\aap},
  574, A14}

\bibitem[{{Cormier} {et~al.}(2015){Cormier}, {Madden}, {Lebouteiller}, {Abel},
  {Hony}, {Galliano}, {R{\'e}my-Ruyer}, {Bigiel}, {Baes}, {Boselli},
  {Chevance}, {Cooray}, {De Looze}, {Doublier}, {Galametz}, {Hughes},
  {Karczewski}, {Lee}, {Lu}, \& {Spinoglio}}]{Cormier:2015}
{Cormier}, D., {Madden}, S.~C., {Lebouteiller}, V., {et~al.} 2015,
  \href{http://dx.doi.org/10.1051/0004-6361/201425207}{\JournalTitle{\aap},
  578, A53}

\bibitem[{{De Looze} {et~al.}(2014){De Looze}, {Cormier}, {Lebouteiller},
  {Madden}, {Baes}, {Bendo}, {Boquien}, {Boselli}, {Clements}, {Cortese},
  {Cooray}, {Galametz}, {Galliano}, {Graci{\'a}-Carpio}, {Isaak}, {Karczewski},
  {Parkin}, {Pellegrini}, {R{\'e}my-Ruyer}, {Spinoglio}, {Smith}, \&
  {Sturm}}]{De-Looze:2014}
{De Looze}, I., {Cormier}, D., {Lebouteiller}, V., {et~al.} 2014,
  \href{http://dx.doi.org/10.1051/0004-6361/201322489}{\JournalTitle{\aap},
  568, A62}

\bibitem[{{De Rosa} {et~al.}(2011){De Rosa}, {Decarli}, {Walter}, {Fan},
  {Jiang}, {Kurk}, {Pasquali}, \& {Rix}}]{De-Rosa:2011}
{De Rosa}, G., {Decarli}, R., {Walter}, F., {et~al.} 2011,
  \href{http://dx.doi.org/10.1088/0004-637X/739/2/56}{\JournalTitle{\apj}, 739,
  56}

\bibitem[{{De Rosa} {et~al.}(2014){De Rosa}, {Venemans}, {Decarli}, {Gennaro},
  {Simcoe}, {Dietrich}, {Peterson}, {Walter}, {Frank}, {McMahon}, {Hewett},
  {Mortlock}, \& {Simpson}}]{De-Rosa:2014}
{De Rosa}, G., {Venemans}, B.~P., {Decarli}, R., {et~al.} 2014,
  \href{http://dx.doi.org/10.1088/0004-637X/790/2/145}{\JournalTitle{\apj},
  790, 145}

\bibitem[{{Decarli} {et~al.}(2017){Decarli}, {Walter}, {Venemans},
  {Ba{\~n}ados}, {Bertoldi}, {Carilli}, {Fan}, {Farina}, {Mazzucchelli},
  {Riechers}, {Rix}, {Strauss}, {Wang}, \& {Yang}}]{Decarli:2017}
{Decarli}, R., {Walter}, F., {Venemans}, B.~P., {et~al.} 2017,
  \href{http://dx.doi.org/10.1038/nature22358}{\JournalTitle{\nat}, 545, 457}

\bibitem[{{D{\'{\i}}az-Santos} {et~al.}(2017){D{\'{\i}}az-Santos}, {Armus},
  {Charmandaris}, {Lu}, {Stierwalt}, {Stacey}, {Malhotra}, {van der Werf},
  {Howell}, {Privon}, {Mazzarella}, {Goldsmith}, {Murphy}, {Barcos-Mu{\~n}oz},
  {Linden}, {Inami}, {Larson}, {Evans}, {Appleton}, {Iwasawa}, {Lord},
  {Sanders}, \& {Surace}}]{Diaz-Santos:2017}
{D{\'{\i}}az-Santos}, T., {Armus}, L., {Charmandaris}, V., {et~al.} 2017,
  \href{http://dx.doi.org/10.3847/1538-4357/aa81d7}{\JournalTitle{\apj}, 846,
  32}

\bibitem[{{Ferrarese} \& {Merritt}(2000)}]{Ferrarese:2000}
{Ferrarese}, L., \& {Merritt}, D. 2000,
  \href{http://dx.doi.org/10.1086/312838}{\JournalTitle{\apjl}, 539, L9}

\bibitem[{{Fisher} {et~al.}(2014){Fisher}, {Bolatto}, {Herrera-Camus},
  {Draine}, {Donaldson}, {Walter}, {Sandstrom}, {Leroy}, {Cannon}, \&
  {Gordon}}]{Fisher:2014}
{Fisher}, D.~B., {Bolatto}, A.~D., {Herrera-Camus}, R., {et~al.} 2014,
  \href{http://dx.doi.org/10.1038/nature12765}{\JournalTitle{\nat}, 505, 186}

\bibitem[{{Goto} {et~al.}(2009){Goto}, {Utsumi}, {Furusawa}, {Miyazaki}, \&
  {Komiyama}}]{Goto:2009}
{Goto}, T., {Utsumi}, Y., {Furusawa}, H., {Miyazaki}, S., \& {Komiyama}, Y.
  2009,
  \href{http://dx.doi.org/10.1111/j.1365-2966.2009.15486.x}{\JournalTitle{\mnras},
  400, 843}

\bibitem[{{Graci{\'a}-Carpio} {et~al.}(2011){Graci{\'a}-Carpio}, {Sturm},
  {Hailey-Dunsheath}, {Fischer}, {Contursi}, {Poglitsch}, {Genzel},
  {Gonz{\'a}lez-Alfonso}, {Sternberg}, {Verma}, {Christopher}, {Davies},
  {Feuchtgruber}, {de Jong}, {Lutz}, \& {Tacconi}}]{Gracia-Carpio:2011}
{Graci{\'a}-Carpio}, J., {Sturm}, E., {Hailey-Dunsheath}, S., {et~al.} 2011,
  \href{http://dx.doi.org/10.1088/2041-8205/728/1/L7}{\JournalTitle{\apjl},
  728, L7}

\bibitem[{{Herrera-Camus} {et~al.}(2015){Herrera-Camus}, {Bolatto}, {Wolfire},
  {Smith}, {Croxall}, {Kennicutt}, {Calzetti}, {Helou}, {Walter}, {Leroy},
  {Draine}, {Brandl}, {Armus}, {Sandstrom}, {Dale}, {Aniano}, {Meidt},
  {Boquien}, {Hunt}, {Galametz}, {Tabatabaei}, {Murphy}, {Appleton}, {Roussel},
  {Engelbracht}, \& {Beirao}}]{Herrera-Camus:2015}
{Herrera-Camus}, R., {Bolatto}, A.~D., {Wolfire}, M.~G., {et~al.} 2015,
  \href{http://dx.doi.org/10.1088/0004-637X/800/1/1}{\JournalTitle{\apj}, 800,
  1}

\bibitem[{{Ho}(2007)}]{Ho:2007}
{Ho}, L.~C. 2007, \href{http://dx.doi.org/10.1086/521917}{\JournalTitle{\apj},
  669, 821}

\bibitem[{{Jiang} {et~al.}(2007){Jiang}, {Fan}, {Vestergaard}, {Kurk},
  {Walter}, {Kelly}, \& {Strauss}}]{Jiang:2007}
{Jiang}, L., {Fan}, X., {Vestergaard}, M., {et~al.} 2007,
  \href{http://dx.doi.org/10.1086/520811}{\JournalTitle{\aj}, 134, 1150}

\bibitem[{{Jones} {et~al.}(2017){Jones}, {Willott}, {Carilli}, {Ferrara},
  {Wang}, \& {Wagg}}]{Jones:2017}
{Jones}, G.~C., {Willott}, C.~J., {Carilli}, C.~L., {et~al.} 2017,
  \href{http://dx.doi.org/10.3847/1538-4357/aa7d0d}{\JournalTitle{\apj}, 845,
  175}

\bibitem[{{Kashikawa} {et~al.}(2015){Kashikawa}, {Ishizaki}, {Willott},
  {Onoue}, {Im}, {Furusawa}, {Toshikawa}, {Ishikawa}, {Niino}, {Shimasaku},
  {Ouchi}, \& {Hibon}}]{Kashikawa:2015}
{Kashikawa}, N., {Ishizaki}, Y., {Willott}, C.~J., {et~al.} 2015,
  \href{http://dx.doi.org/10.1088/0004-637X/798/1/28}{\JournalTitle{\apj}, 798,
  28}

\bibitem[{{Katz} {et~al.}(2017){Katz}, {Kimm}, {Sijacki}, \&
  {Haehnelt}}]{Katz:2017}
{Katz}, H., {Kimm}, T., {Sijacki}, D., \& {Haehnelt}, M.~G. 2017,
  \href{http://dx.doi.org/10.1093/mnras/stx608}{\JournalTitle{\mnras}, 468,
  4831}

\bibitem[{{Khandai} {et~al.}(2015){Khandai}, {Di Matteo}, {Croft}, {Wilkins},
  {Feng}, {Tucker}, {DeGraf}, \& {Liu}}]{Khandai:2015}
{Khandai}, N., {Di Matteo}, T., {Croft}, R., {et~al.} 2015,
  \href{http://dx.doi.org/10.1093/mnras/stv627}{\JournalTitle{\mnras}, 450,
  1349}

\bibitem[{{Kormendy} \& {Ho}(2013)}]{Kormendy:2013}
{Kormendy}, J., \& {Ho}, L.~C. 2013,
  \href{http://dx.doi.org/10.1146/annurev-astro-082708-101811}{\JournalTitle{\araa},
  51, 511}

\bibitem[{{Lauer} {et~al.}(2007){Lauer}, {Tremaine}, {Richstone}, \&
  {Faber}}]{Lauer:2007}
{Lauer}, T.~R., {Tremaine}, S., {Richstone}, D., \& {Faber}, S.~M. 2007,
  \href{http://dx.doi.org/10.1086/522083}{\JournalTitle{\apj}, 670, 249}

\bibitem[{{Maiolino} {et~al.}(2005){Maiolino}, {Cox}, {Caselli}, {Beelen},
  {Bertoldi}, {Carilli}, {Kaufman}, {Menten}, {Nagao}, {Omont}, {Wei{\ss}},
  {Walmsley}, \& {Walter}}]{Maiolino:2005}
{Maiolino}, R., {Cox}, P., {Caselli}, P., {et~al.} 2005,
  \href{http://dx.doi.org/10.1051/0004-6361:200500165}{\JournalTitle{\aap},
  440, L51}

\bibitem[{{Maiolino} {et~al.}(2015){Maiolino}, {Carniani}, {Fontana},
  {Vallini}, {Pentericci}, {Ferrara}, {Vanzella}, {Grazian}, {Gallerani},
  {Castellano}, {Cristiani}, {Brammer}, {Santini}, {Wagg}, \&
  {Williams}}]{Maiolino:2015a}
{Maiolino}, R., {Carniani}, S., {Fontana}, A., {et~al.} 2015,
  \href{http://dx.doi.org/10.1093/mnras/stv1194}{\JournalTitle{\mnras}, 452,
  54}

\bibitem[{{Narayanan} \& {Krumholz}(2017)}]{Narayanan:2017}
{Narayanan}, D., \& {Krumholz}, M.~R. 2017,
  \href{http://dx.doi.org/10.1093/mnras/stw3218}{\JournalTitle{\mnras}, 467,
  50}

\bibitem[{{Omont} {et~al.}(2013){Omont}, {Willott}, {Beelen}, {Bergeron},
  {Orellana}, \& {Delorme}}]{Omont:2013}
{Omont}, A., {Willott}, C.~J., {Beelen}, A., {et~al.} 2013,
  \href{http://dx.doi.org/10.1051/0004-6361/201221006}{\JournalTitle{\aap},
  552, A43}

\bibitem[{{Pallottini} {et~al.}(2017){Pallottini}, {Ferrara}, {Gallerani},
  {Vallini}, {Maiolino}, \& {Salvadori}}]{Pallottini:2017}
{Pallottini}, A., {Ferrara}, A., {Gallerani}, S., {et~al.} 2017,
  \href{http://dx.doi.org/10.1093/mnras/stw2847}{\JournalTitle{\mnras}, 465,
  2540}

\bibitem[{{Planck Collaboration}(2014)}]{Planck-Collaboration:2014}
{Planck Collaboration}. 2014,
  \href{http://dx.doi.org/10.1051/0004-6361/201321591}{\JournalTitle{\aap},
  571, A16}

\bibitem[{{Robson} {et~al.}(2004){Robson}, {Priddey}, {Isaak}, \&
  {McMahon}}]{Robson:2004}
{Robson}, I., {Priddey}, R.~S., {Isaak}, K.~G., \& {McMahon}, R.~G. 2004,
  \href{http://dx.doi.org/10.1111/j.1365-2966.2004.07923.x}{\JournalTitle{\mnras},
  351, L29}

\bibitem[{{Sargsyan} {et~al.}(2014){Sargsyan}, {Samsonyan}, {Lebouteiller},
  {Weedman}, {Barry}, {Bernard-Salas}, {Houck}, \& {Spoon}}]{Sargsyan:2014}
{Sargsyan}, L., {Samsonyan}, A., {Lebouteiller}, V., {et~al.} 2014,
  \href{http://dx.doi.org/10.1088/0004-637X/790/1/15}{\JournalTitle{\apj}, 790,
  15}

\bibitem[{{Sargsyan} {et~al.}(2012){Sargsyan}, {Lebouteiller}, {Weedman},
  {Spoon}, {Bernard-Salas}, {Engels}, {Stacey}, {Houck}, {Barry}, {Miles}, \&
  {Samsonyan}}]{Sargsyan:2012}
{Sargsyan}, L., {Lebouteiller}, V., {Weedman}, D., {et~al.} 2012,
  \href{http://dx.doi.org/10.1088/0004-637X/755/2/171}{\JournalTitle{\apj},
  755, 171}

\bibitem[{{Shao} {et~al.}(2017){Shao}, {Wang}, {Jones}, {Carilli}, {Walter},
  {Fan}, {Riechers}, {Bertoldi}, {Wagg}, {Strauss}, {Omont}, {Cox}, {Jiang},
  {Narayanan}, \& {Menten}}]{Shao:2017}
{Shao}, Y., {Wang}, R., {Jones}, G.~C., {et~al.} 2017,
  \href{http://dx.doi.org/10.3847/1538-4357/aa826c}{\JournalTitle{\apj}, 845,
  138}

\bibitem[{{Smith} {et~al.}(2017){Smith}, {Croxall}, {Draine}, {De Looze},
  {Sandstrom}, {Armus}, {Beir{\~a}o}, {Bolatto}, {Boquien}, {Brandl},
  {Crocker}, {Dale}, {Galametz}, {Groves}, {Helou}, {Herrera-Camus}, {Hunt},
  {Kennicutt}, {Walter}, \& {Wolfire}}]{Smith:2017}
{Smith}, J.~D.~T., {Croxall}, K., {Draine}, B., {et~al.} 2017,
  \href{http://dx.doi.org/10.3847/1538-4357/834/1/5}{\JournalTitle{\apj}, 834,
  5}

\bibitem[{{Stacey} {et~al.}(2010){Stacey}, {Hailey-Dunsheath}, {Ferkinhoff},
  {Nikola}, {Parshley}, {Benford}, {Staguhn}, \& {Fiolet}}]{Stacey:2010}
{Stacey}, G.~J., {Hailey-Dunsheath}, S., {Ferkinhoff}, C., {et~al.} 2010,
  \href{http://dx.doi.org/10.1088/0004-637X/724/2/957}{\JournalTitle{\apj},
  724, 957}

\bibitem[{{Stefan} {et~al.}(2015){Stefan}, {Carilli}, {Wagg}, {Walter},
  {Riechers}, {Bertoldi}, {Green}, {Fan}, {Menten}, \& {Wang}}]{Stefan:2015}
{Stefan}, I.~I., {Carilli}, C.~L., {Wagg}, J., {et~al.} 2015,
  \href{http://dx.doi.org/10.1093/mnras/stv1108}{\JournalTitle{\mnras}, 451,
  1713}

\bibitem[{{Venemans} {et~al.}(2016){Venemans}, {Walter}, {Zschaechner},
  {Decarli}, {De Rosa}, {Findlay}, {McMahon}, \& {Sutherland}}]{Venemans:2016}
{Venemans}, B.~P., {Walter}, F., {Zschaechner}, L., {et~al.} 2016,
  \href{http://dx.doi.org/10.3847/0004-637X/816/1/37}{\JournalTitle{\apj}, 816,
  37}

\bibitem[{{Venemans} {et~al.}(2012){Venemans}, {McMahon}, {Walter}, {Decarli},
  {Cox}, {Neri}, {Hewett}, {Mortlock}, {Simpson}, \& {Warren}}]{Venemans:2012}
{Venemans}, B.~P., {McMahon}, R.~G., {Walter}, F., {et~al.} 2012,
  \href{http://dx.doi.org/10.1088/2041-8205/751/2/L25}{\JournalTitle{\apjl},
  751, L25}

\bibitem[{{Venemans} {et~al.}(2013){Venemans}, {Findlay}, {Sutherland}, {De
  Rosa}, {McMahon}, {Simcoe}, {Gonz{\'a}lez-Solares}, {Kuijken}, \&
  {Lewis}}]{Venemans:2013}
{Venemans}, B.~P., {Findlay}, J.~R., {Sutherland}, W.~J., {et~al.} 2013,
  \href{http://dx.doi.org/10.1088/0004-637X/779/1/24}{\JournalTitle{\apj}, 779,
  24}

\bibitem[{{Venemans} {et~al.}(2015){Venemans}, {Ba{\~n}ados}, {Decarli},
  {Farina}, {Walter}, {Chambers}, {Fan}, {Rix}, {Schlafly}, {McMahon},
  {Simcoe}, {Stern}, {Burgett}, {Draper}, {Flewelling}, {Hodapp}, {Kaiser},
  {Magnier}, {Metcalfe}, {Morgan}, {Price}, {Tonry}, {Waters}, {AlSayyad},
  {Banerji}, {Chen}, {Gonz{\'a}lez-Solares}, {Greiner}, {Mazzucchelli},
  {McGreer}, {Miller}, {Reed}, \& {Sullivan}}]{Venemans:2015a}
{Venemans}, B.~P., {Ba{\~n}ados}, E., {Decarli}, R., {et~al.} 2015,
  \href{http://dx.doi.org/10.1088/2041-8205/801/1/L11}{\JournalTitle{\apjl},
  801, L11}

\bibitem[{{Venemans} {et~al.}(2017){Venemans}, {Walter}, {Decarli},
  {Ba{\~n}ados}, {Hodge}, {Hewett}, {McMahon}, {Mortlock}, \&
  {Simpson}}]{Venemans:2017}
{Venemans}, B.~P., {Walter}, F., {Decarli}, R., {et~al.} 2017,
  \href{http://dx.doi.org/10.3847/1538-4357/aa62ac}{\JournalTitle{\apj}, 837,
  146}

\bibitem[{{Volonteri} {et~al.}(2015){Volonteri}, {Capelo}, {Netzer},
  {Bellovary}, {Dotti}, \& {Governato}}]{Volonteri:2015}
{Volonteri}, M., {Capelo}, P.~R., {Netzer}, H., {et~al.} 2015,
  \href{http://dx.doi.org/10.1093/mnras/stv387}{\JournalTitle{\mnras}, 449,
  1470}

\bibitem[{{Walter} {et~al.}(2004){Walter}, {Carilli}, {Bertoldi}, {Menten},
  {Cox}, {Lo}, {Fan}, \& {Strauss}}]{Walter:2004}
{Walter}, F., {Carilli}, C., {Bertoldi}, F., {et~al.} 2004,
  \href{http://dx.doi.org/10.1086/426017}{\JournalTitle{\apjl}, 615, L17}

\bibitem[{{Wang} {et~al.}(2008){Wang}, {Wagg}, {Carilli}, {Benford}, {Dowell},
  {Bertoldi}, {Walter}, {Menten}, {Omont}, {Cox}, {Strauss}, {Fan}, \&
  {Jiang}}]{Wang:2008}
{Wang}, R., {Wagg}, J., {Carilli}, C.~L., {et~al.} 2008,
  \href{http://dx.doi.org/10.1088/0004-6256/135/4/1201}{\JournalTitle{\aj},
  135, 1201}

\bibitem[{{Wang} {et~al.}(2011){Wang}, {Wagg}, {Carilli}, {Neri}, {Walter},
  {Omont}, {Riechers}, {Bertoldi}, {Menten}, {Cox}, {Strauss}, {Fan}, \&
  {Jiang}}]{Wang:2011}
---. 2011,
  \href{http://dx.doi.org/10.1088/0004-6256/142/4/101}{\JournalTitle{\aj}, 142,
  101}

\bibitem[{{Wang} {et~al.}(2013){Wang}, {Wagg}, {Carilli}, {Walter}, {Lentati},
  {Fan}, {Riechers}, {Bertoldi}, {Narayanan}, {Strauss}, {Cox}, {Omont},
  {Menten}, {Knudsen}, {Neri}, \& {Jiang}}]{Wang:2013}
---. 2013,
  \href{http://dx.doi.org/10.1088/0004-637X/773/1/44}{\JournalTitle{\apj}, 773,
  44}

\bibitem[{{Wang} {et~al.}(2016){Wang}, {Wu}, {Neri}, {Fan}, {Walter},
  {Carilli}, {Momjian}, {Bertoldi}, {Strauss}, {Li}, {Wang}, {Riechers},
  {Jiang}, {Omont}, {Wagg}, \& {Cox}}]{Wang:2016}
{Wang}, R., {Wu}, X.-B., {Neri}, R., {et~al.} 2016,
  \href{http://dx.doi.org/10.3847/0004-637X/830/1/53}{\JournalTitle{\apj}, 830,
  53}

\bibitem[{{Willott} {et~al.}(2015{\natexlab{a}}){Willott}, {Bergeron}, \&
  {Omont}}]{Willott:2015}
{Willott}, C.~J., {Bergeron}, J., \& {Omont}, A. 2015{\natexlab{a}},
  \href{http://dx.doi.org/10.1088/0004-637X/801/2/123}{\JournalTitle{\apj},
  801, 123}

\bibitem[{{Willott} {et~al.}(2015{\natexlab{b}}){Willott}, {Carilli}, {Wagg},
  \& {Wang}}]{Willott:2015a}
{Willott}, C.~J., {Carilli}, C.~L., {Wagg}, J., \& {Wang}, R.
  2015{\natexlab{b}},
  \href{http://dx.doi.org/10.1088/0004-637X/807/2/180}{\JournalTitle{\apj},
  807, 180}

\bibitem[{{Willott} {et~al.}(2011){Willott}, {Chet}, {Bergeron}, \&
  {Hutchings}}]{Willott:2011}
{Willott}, C.~J., {Chet}, S., {Bergeron}, J., \& {Hutchings}, J.~B. 2011,
  \href{http://dx.doi.org/10.1088/0004-6256/142/6/186}{\JournalTitle{\aj}, 142,
  186}

\bibitem[{{Willott} {et~al.}(2013){Willott}, {Omont}, \&
  {Bergeron}}]{Willott:2013}
{Willott}, C.~J., {Omont}, A., \& {Bergeron}, J. 2013,
  \href{http://dx.doi.org/10.1088/0004-637X/770/1/13}{\JournalTitle{\apj}, 770,
  13}

\bibitem[{{Willott} {et~al.}(2005){Willott}, {Percival}, {McLure}, {Crampton},
  {Hutchings}, {Jarvis}, {Sawicki}, \& {Simard}}]{Willott:2005a}
{Willott}, C.~J., {Percival}, W.~J., {McLure}, R.~J., {et~al.} 2005,
  \href{http://dx.doi.org/10.1086/430168}{\JournalTitle{\apj}, 626, 657}

\bibitem[{{Willott} {et~al.}(2007){Willott}, {Delorme}, {Omont}, {Bergeron},
  {Delfosse}, {Forveille}, {Albert}, {Reyl{\'e}}, {Hill}, {Gully-Santiago},
  {Vinten}, {Crampton}, {Hutchings}, {Schade}, {Simard}, {Sawicki}, {Beelen},
  \& {Cox}}]{Willott:2007a}
{Willott}, C.~J., {Delorme}, P., {Omont}, A., {et~al.} 2007,
  \href{http://dx.doi.org/10.1086/522962}{\JournalTitle{\aj}, 134, 2435}

\bibitem[{{Willott} {et~al.}(2009){Willott}, {Delorme}, {Reyl{\'e}}, {Albert},
  {Bergeron}, {Crampton}, {Delfosse}, {Forveille}, {Hutchings}, {McLure},
  {Omont}, \& {Schade}}]{Willott:2009}
{Willott}, C.~J., {Delorme}, P., {Reyl{\'e}}, C., {et~al.} 2009,
  \href{http://dx.doi.org/10.1088/0004-6256/137/3/3541}{\JournalTitle{\aj},
  137, 3541}

\bibitem[{{Willott} {et~al.}(2010{\natexlab{a}}){Willott}, {Albert},
  {Arzoumanian}, {Bergeron}, {Crampton}, {Delorme}, {Hutchings}, {Omont},
  {Reyl{\'e}}, \& {Schade}}]{Willott:2010}
{Willott}, C.~J., {Albert}, L., {Arzoumanian}, D., {et~al.} 2010{\natexlab{a}},
  \href{http://dx.doi.org/10.1088/0004-6256/140/2/546}{\JournalTitle{\aj}, 140,
  546}

\bibitem[{{Willott} {et~al.}(2010{\natexlab{b}}){Willott}, {Delorme},
  {Reyl{\'e}}, {Albert}, {Bergeron}, {Crampton}, {Delfosse}, {Forveille},
  {Hutchings}, {McLure}, {Omont}, \& {Schade}}]{Willott:2010a}
{Willott}, C.~J., {Delorme}, P., {Reyl{\'e}}, C., {et~al.} 2010{\natexlab{b}},
  \href{http://dx.doi.org/10.1088/0004-6256/139/3/906}{\JournalTitle{\aj}, 139,
  906}

\bibitem[{{Wu} {et~al.}(2015){Wu}, {Wang}, {Fan}, {Yi}, {Zuo}, {Bian}, {Jiang},
  {McGreer}, {Wang}, {Yang}, {Yang}, {Thompson}, \& {Beletsky}}]{Wu:2015}
{Wu}, X.-B., {Wang}, F., {Fan}, X., {et~al.} 2015,
  \href{http://dx.doi.org/10.1038/nature14241}{\JournalTitle{\nat}, 518, 512}

\end{thebibliography}

\end{document}